\documentclass[english,pra,amsfonts,amssymb,amsmath,twocolumn,groupedaddress]{revtex4-1}
\usepackage[utf8]{inputenc}
\pdfpageattr{/Group <</S /Transparency /I true /CS /DeviceRGB>>}
\usepackage{amsthm}
\usepackage{amsmath}
\usepackage{amssymb}
\usepackage{subcaption}

\usepackage{tikz} 

\PassOptionsToPackage{normalem}{ulem}
\usepackage{ulem}

\usepackage{comment}
\usepackage{caption}
\usepackage{graphicx}

\usepackage{algpseudocode}

\algnewcommand{\A}{\textbf{and}\space}
\algnewcommand{\Or}{\textbf{or}\space}
\algnewcommand{\Xor}{\textbf{xor}\space}
\makeatletter
\let\OldStatex\Statex
\renewcommand{\Statex}[1][3]{%
  \setlength\@tempdima{\algorithmicindent}%
  \OldStatex\hskip\dimexpr#1\@tempdima\relax}
\makeatother

\makeatletter

\theoremstyle{plain}
\newtheorem{thm}{\protect\theoremname}
\theoremstyle{plain}

\newtheorem{lem}[thm]{Lemma}
\newtheorem{defin}[thm]{\protect\definname}
\newtheorem{observ}[thm]{\protect\observname}
\newtheorem{corol}[thm]{\protect\corolname}

\newtheorem{cnj}[thm]{Conjecture}
\newtheorem{example}[thm]{\protect\examplename}

\makeatother

\usepackage{babel}

\providecommand{\theoremname}{Theorem}
\providecommand{\definname}{Definition}
\providecommand{\observname}{Observation}
\providecommand{\corolname}{Corollary}
\providecommand{\algorithmname}{Algorithm}
\providecommand{\examplename}{Example}
\providecommand{\problemname}{Problem}

\usepackage{pgfplots}
\usetikzlibrary{calc}
\usepackage{pgfplotstable}
\usepackage{colortbl}
\usepackage{booktabs}
\usepackage{pgfplotstable}

\newcommand{\ket}[1]{|#1\rangle}

\newcommand{\onemat}[0]{{\mathbf 1}}

\newcommand{\nix}[1]{{}}

\begin{document}

\title{Efficient Synthesis of Probabilistic Quantum Circuits with Fallback}

\author{Alex Bocharov$^*$}
\author{Martin Roetteler$^*$}
\author{Krysta M.~Svore$^*$}

\affiliation{
$^*$Quantum Architectures and Computation Group, Microsoft Research, Redmond, WA (USA)}

\begin{abstract}

Recently it has been shown that Repeat-Until-Success (RUS) circuits can approximate a given single-qubit unitary with an expected number of $T$ gates of about $1/3$ of what is required by optimal, deterministic, ancilla-free decompositions over the Clifford+$T$ gate set.
In this work, we introduce a more general and conceptually simpler circuit decomposition method that allows for synthesis into protocols that probabilistically implement quantum circuits over several universal gate sets including, but not restricted to, the Clifford+$T$ gate set.  The protocol, which we call  Probabilistic Quantum Circuits with Fallback (PQF), implements a walk on a discrete Markov chain in which the target unitary is an absorbing state and in which transitions are induced by multi-qubit unitaries followed by measurements. In contrast to RUS protocols, the presented PQF protocols terminate after a finite number of steps. Specifically, we apply our method to the Clifford+$T$, Clifford+$V$, and Clifford+$\pi/12$ gate sets to achieve decompositions with expected gate counts of $\log_b(1/\varepsilon)+O(\log(\log(1/\varepsilon)))$, where $b$ is a quantity related to the expansion property of the underlying universal gate set.

\end{abstract}

\maketitle

\section{Introduction}

Techniques to efficiently compile higher-level quantum algorithms into lower-level fault-tolerant circuits are a critical step for the implementation of a scalable, general purpose quantum computer. Several universal fault-tolerant gate sets arise from augmenting the set of Clifford gates by additional gates that arise naturally from the underlying fault-tolerance scheme. An important example is the Clifford+$T$ basis, consisting of controlled-NOTs ($\mbox{CNOT}$) and Hadamard ($H$) gates, together with the $T$ gate, which is given by
$T = \left[\begin{smallmatrix}1&0\\0&e^{i \pi/4}\end{smallmatrix}\right]$.
Further examples of interest are the Clifford$+V$ basis in which the set of Clifford gates is augmented by the $6$ non-Clifford gates $\frac{1}{\sqrt{5}}(1\pm 2iP)$, where $P\in \{X, Y, Z\}$, and the Clifford$+\pi/12$ basis in which the gate $K = \left[\begin{smallmatrix}1&0\\0&e^{i \pi/6}\end{smallmatrix}\right]$ is added. 

While the Solovay-Kitaev algorithm \cite{IkeAndMike2000,DN} allows to solve the synthesis problem for any universal gate set, there are certain disadvantages to this approach, in particular the large depth of the resulting circuits: to  the best of our knowledge the resulting depth is only known to scale as $O(\log^{3.97} (1/\varepsilon))$, where $\varepsilon$ is the target approximation error with which the single qubit unitary has to be implemented. Also the compilation-time of the Solovay-Kitaev method, i.e., the time it takes to execute the classical algorithm that produces the output circuit is quite high, namely almost cubic in $\log(1/\varepsilon)$. This makes the application of the algorithm for small values of the target precision, say in a regime where $\varepsilon \sim 10^{-15}$, difficult if not impossible. On the other hand, there exist several quantum algorithms that would require this level of target approximation error in order to scale to instance sizes of practical interest. 

Happily, it was shown recently~\cite{Selinger,RoSelinger,KMM1231,Kliuchnikoff} that for the Clifford$+T$ basis, elementary number theory can be leveraged to obtain much more efficient algorithms for approximating a single-qubit gates. We refer to these methods as being deterministic and ancilla-free as they lead to a decomposition of the target unitary that can be executed in an entirely pre-determined sequence of single-qubit unitaries over the given gate set.
The number of $T$ gates in the resulting circuits scales close to $3\log_2(1/\varepsilon)$
for $Z$-rotations, which is within a constant factor of the information-theoretic lower bound. Also, the compilation-time of these methods is low: using reasonable number-theoretic conjectures for which there exists an overwhelming amount of empirical evidence, the compilation-time follows the same scaling (up to logarithmic factors). Any non-axial rotation $V$ can be decomposed into axial rotations such that
\cite{IkeAndMike2000}
\begin{equation}\label{eq:Euler}
V = e^{i \delta}   R_z(\alpha)   H   R_z(\beta)   H   R_z(\gamma),
\end{equation}
for real values $\alpha, \beta, \gamma, \delta$. This yields an upper bound of 
$9\log_2(1/\varepsilon)$ for all of the above mentioned deterministic, ancilla-free methods for general rotations.  

In contrast to this, it was recently shown \cite{BSRRUS} that by using non-deterministic circuits that employ a small number of ancilla qubits, the number of $T$ gates can be further reduced by a factor of $2.5$ on average for axial rotations, namely to $O(1.15 \log_2(1/\varepsilon))$. Again, using Euler angles, this leads to a complexity of an expected number of $O(3.45 \log_2(1/\varepsilon))$ for general rotations. For decomposition of a given unitary $U$, these so-called Repeat-Until-Success (RUS) circuits \cite{PS} consist of repeated application of a Clifford+$T$ sequence on an input state $\ket{\psi}$ and an ancilla qubit, followed by measurement of the ancilla qubit to project the input state $\ket{\psi}$ to the state $U\ket{\psi}$ \cite{PS,BSRRUS}.

An RUS circuit allows for a potentially unlimited sequence of trial and correction cycles with guaranteed finite {\em expected cost} below the lower bound achieved by a purely unitary circuit design.
The correction circuit in each cycle can be designed to have zero cost, namely by requiring it to be a circuit consisting only of Pauli gates.
The synthesis algorithm for RUS circuits over the Clifford+$T$ basis is based on a randomized search and achieves an expected mean $T$-count with a leading term of $(1+\delta)\log_2(1/\varepsilon)$, where $\delta=0.15$ was achievable for practically important precisions $\varepsilon$ \cite{BSRRUS}.

In contrast, the Probabilistic Quantum Circuit with Fallback (PQF) protocols introduced in the present work entail at most a small finite number of trials (possibly all different), and one final, purely unitary, correction step.
The final correction step, or \emph{fallback}, may have considerable cost, however the probability of requiring the fallback step can be very small allowing for an improved expected cost for decomposition.
Synthesizing a PQF circuit to approximate a given target is remarkably simpler than in the RUS case.
In addition, we generalize PQF to three universal quantum bases: Clifford+$T$, Clifford+$V$ \cite{GENUINE}, and Clifford+$\pi/12$.
Clifford+$V$ was previously considered for purely unitary, deterministic decomposition of single-qubit gates and resulted in the shortest known single-qubit circuits \cite{BGV,Ross}.
Clifford+$\pi/12$ has been identified as relevant for quantum computer architectures based on metaplectic anyons \cite{Metaplectic}.

We present an efficient algorithm for single-qubit decomposition based on our PQF protocol.
We describe the algorithmic steps in detail for each of the three bases considered.
Our algorithm achieves an expected gate count of $\log_b(1/\varepsilon)+O(\log(\log(1/\varepsilon)))$, where $b$ is related to the scaling of the number of unique circuits that can be formed over the underlying basis. More precisely, $b$ is defined so that for a given depth $t$ the number of unique circuits scales as $\Theta(b^t)$, i.e., $b$ characterizes the expansion of the underlying set of generators. 
Specifically, we have $b=2$ for Clifford+$T$, $b=5$ for Clifford+$V$, and $b=4$ for Clifford+$\pi/12$.
The PQF protocol can be generalized to several bases as the exactly representable unitaries over a given basis are representable as unitarizations of matrices over rings of cyclotomic integers of order $4m$, $m \in \{1,2,3\}$, where $m=1$ for Clifford+$V$, $2$ for Clifford+$T$, and $3$ for Clifford+$\pi/12$.
We aim to generalize our designs to other cyclotomic orders in future work.



\section{Design of Probabilistic Quantum Circuits with Fallback}

In this section, we define Probabilistic Quantum Circuits with Fallback (PQF).
Our PQF protocol employs both probabilistic and deterministic subcircuits.  The former are referred to as \emph{primary} and the latter as \emph{fallback}.
We focus on the case of PQF circuits for single qubit unitaries that are axial rotations around the $Z$-axis, which by the Euler angle decomposition Eq~(\ref{eq:Euler}) will imply PQF protocols for arbitary single qubit unitaries. However, we point out that in principle the probabilistic circuit design described in this paper can also be applied to multi-qubit unitaries and even to systems consisting of higher-dimensional subsystems, such as e.g. qutrits.  

The primary subcircuits can be synthesized using existing synthesis methods \cite{KMM12,GENUINE} that given $G$ and $\varepsilon$ generate a probabilistic circuit $P(G,\varepsilon)$ to perform an $\varepsilon$-approximation of the gate $G$ with probability $p>0$ and performs some other unitary gate $G_1$ with probability $1-p$.
Let $C_P(G, \varepsilon)$ be the execution cost of the $P(G,\varepsilon)$ circuit.
%
The \emph{fallback} subcircuit can be constructed using a synthesis method that for a given unitary target gate $G$ and a desired precision $\varepsilon$ generates an $\varepsilon$-approximation circuit $F(G,\varepsilon)$ with a known execution cost $C_F(G,\varepsilon)$, such as those in \cite{RoSelinger,Kliuchnikoff,GENUINE}.

If $C_P(G,\varepsilon)$ is uniformly smaller than $p   C_F(G,\varepsilon)$, then an $\varepsilon$-approximation of the gate $G$ using PQF will have lower expected cost than implementing the fallback circuit at cost $C_F(G,\varepsilon)$.
To this end, we create a circuit with classical feedback that first performs the subcircuit $P(G, \varepsilon)$ with the (desired) outcome $\sim G \ket{\psi}$ upon measuring 0 or (undesired) outcome $G_1 \ket{\psi}$ upon measuring 1.
If the measurement outcome is 1, the circuit then performs $F(G G_1^{\dagger}, \varepsilon)$ on $G_1\ket{\psi}$.
The expected cost the entire circuit protocol is $C_P(G,\varepsilon)+(1-p) C_F(G G_1^{\dagger},\varepsilon)$, which is smaller than $C_F(G, \varepsilon)$ if and only if $C_P(G,\varepsilon)< p  C_F(G G_1^{\dagger},\varepsilon)$.

The concatenated circuit, denoted as $PQF(G, \varepsilon, 1)$, is a special case of a nested probabilistic circuit,
$PQF(G,\varepsilon,k)$, $k \in \mathbb{Z}$, $k \geq 0$, defined inductively as follows:
\begin{eqnarray}
PQF(G, \varepsilon, 0) &=& F(G, \varepsilon)\\ \nonumber
PQF(G,\varepsilon,k) &=& \\ \nonumber
&&P(G,\varepsilon) \cup   BC(PQF(G F_k^{\dagger},\varepsilon,k-1)),
\end{eqnarray}
where $F_k$ is the undesirable outcome of the $P$ circuit and $BC$ denotes binary classical control on such an outcome.

\begin{figure}[tb]
\centering
\includegraphics[width=3.5in]{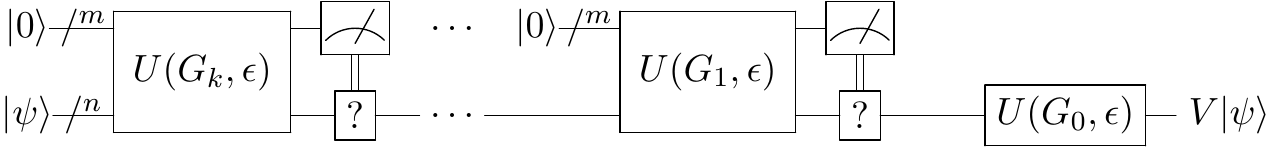}
\caption{\label{fig:pqcf-circuit}
PQF protocol to implement unitary $V=G_k$.
}
\end{figure}

The general layout of a PQF circuit is shown in Figure \ref{fig:pqcf-circuit}.
The ``question mark" box denotes the binary classical control switch that implements the remainder of the circuit if and only if the measurement result is 1.
Let $F_{j} |\psi\rangle$ be the undesired result upon measurement of 1 at the $j$-th round of the protocol.
Then $G_{j-1}=G_j   F_j^{\dagger}$ and we note that the synthesis algorithm computes $\Pi_{j=0}^{k-1} G_j$.

It follows that if $PQF(G,\varepsilon,1)$ is a cost improvement over $PQF(G,\varepsilon,0)$, then $PQF(G,\varepsilon,k)$ is a cost improvement over $PQF(G,\varepsilon,k-1)$ for any $k>0$.
However, we show that the incremental improvement scales like $O((1-p)^k)$ and therefore near-optimal performance can be achieved with a relatively small number of rounds $k$.

Both the PQF and RUS protocol require synthesis of unitary subcircuits.
In PQF, the primary probabilistic subcircuits vary at each round, and the final fallback circuit, if necessary, is deterministic.
In RUS, the same probabilistic subcircuit is applied in each round, followed by the same correction if necessary.

%
%

\begin{figure*}[hbt]

\tikzstyle{state}=[circle,thick,minimum size=1cm,draw=black!80,fill=black!10]

\begin{tabular}{ll}
\begin{tikzpicture}[>=latex,text height=1.5ex,text depth=0.25ex]
   \matrix[row sep=1cm,column sep=1cm] {
     \node (i_1) [state]{$I$}; &
     \node (i_2) [state]{$I$}; &
     \node (i_3) [state]{$I$}; &
     \node (i_4) {$\cdots$}; &
     \\
     & \node (j_2) [state]{$G$}; &
     \\
   };
   \path[->]
     (i_1) edge[thick] node [above] {$1{-}p$} (i_2)
     (i_2) edge[thick] node [above] {$1{-}p$} (i_3)
     (i_3) edge[thick] node [above] {$1{-}p$} (i_4)
     (i_1) edge[thick] node [left] {$p$\;}(j_2)
     (i_2) edge[thick] node [left] {$p$\;}(j_2)
     (i_3) edge[thick] node [left] {$p$\;} (j_2)
    ;
\end{tikzpicture}
&
\begin{tikzpicture}[>=latex,text height=1.5ex,text depth=0.25ex]
   \matrix[row sep=1cm,column sep=1.2cm] {
     \node (i_1) [state]{$I$}; &
     \node (i_2) [state]{$F_k$}; &
     \node (i_3) [state]{$F_{k-1}$}; &
     \node (i_4) {$\cdots$}; &
     \node (i_5) [state]{$F_1$}; &
     \\
     & \node (j_2) [state]{$G$}; &
     \\
   };
   \path[->]
     (i_1) edge[thick] node [above] {$1{-}p_k$} (i_2)
     (i_2) edge[thick] node [above] {$1{-}p_{k-1}$} (i_3)
     (i_3) edge[thick] node [above] {$1{-}p_{k-2}$} (i_4)
     (i_4) edge[thick] (i_5)
     (i_1) edge[thick] node [left] {$p_k$\;}(j_2)
     (i_2) edge[thick] node [left] {$p_{k-1}$\;}(j_2)
     (i_3) edge[thick] node [left] {$p_{k-2}$\;} (j_2)
     (i_5) edge[thick] node [left] {$1$ \;\;\; } (j_2)
    ;
\end{tikzpicture}
\\
a) & b)
\end{tabular}
\caption{\label{fig:markov} Markov chains for the implementation of a target unitary transformation $G$. Shown are state transitions for a) Repeat-Until-Success (RUS) protocols and b) Probabilistic Circuits with Fallback (PQF) protocols.}
\end{figure*}
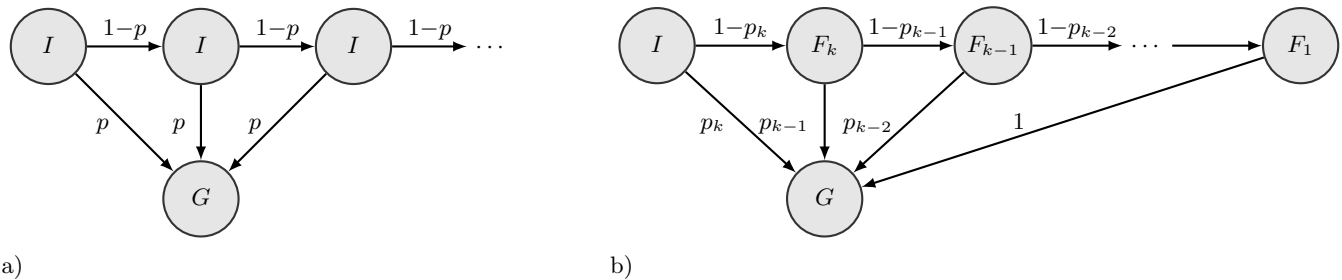

In Figure \ref{fig:markov} we compare the Markov chain \cite{Norris:98} corresponding to the implementation of an RUS protocol to that corresponding to  a PQF protocol. Both protocols implement a target single qubit unitary transformation $G$ for a given target approximation $\varepsilon$ by performing a random walk on the nodes, where the target unitary $G$ is an absorbing state, i.e., the walk terminates once it arrives in this state. In general, each node represents the unitary transformation that has been applied to the input state at the respective stage in the protocol. Both, in case of RUS and PQF protocols, each transition between nodes is probabilistic and is induced by the success or failure of a multi-qubit unitary followed by a measurement. In the case of the RUS protocol shown in Figure \ref{fig:markov} a), the applied transformation in case of failure is always the identity---or, more generally, a local Clifford operation which however can easily be corrected to become the identity, whence we represent this case by the identity operator $I$ also---whereas in case of the PQF protocol each intermediate node corresponds to an operation $F_j$, i.e., as shown in Figure~\ref{fig:pqcf-circuit} b), the state is $F_j \ket{\psi_j}$, where $F_{j}$ be the undesired result upon measurement of 1 at the $j$-th attempt of the protocol and $\ket{\psi_j}$ was the state in the previous round, where $j=k,k-1, \ldots, 1$. The probability of success of this step is denoted by $p_j$ and the probability of failure correspondingly by $1-p_j$. 

If we find ourselves in node $j$ of the protocol we not only know the entire history of previous failed attempts to implement the target gate $G$, we can also attempt to reach the target state $G$ by applying a probabilistic circuit 
that implements $G_{j-1} := G_j F_j^{\dagger}$ where $G_j$ was defined in a previous round. In any case we will implement $G$ after at most $k$ steps as $G = \Pi_{j=0}^{\ell} G_j$, where $\ell \in \{0,\ldots,k\}$ denotes the first point in time where the protocol had a successful transition to the target gate $G$. 

It is useful to think of the probabilistic transitions into the nodes labeled with $F_j$ for $j=k, k-1, \ldots, 2$ as being very cheap, whereas the last transition (the ``fallback'') from $F_1 \rightarrow G$ is expensive, but will always lead to the absorbing state, i.e., it guarantees that the protocol implements $G$ with precision $\varepsilon$ after at most $k$ rounds. 

Note the that RUS designs might require a potentially unbounded number of iterations to reach the accepting state to implement the target gate $G$. In contrast, a PQF design with $k$ stages is guaranteed to always implement the target gate after at most $k$ attempts. An important difference between the two models is that in (a) the target for each transition is independent of the stage so that the cost just depends on the target unitary $G$ and the target error $\varepsilon$, whereas in (b) the target depends on the stage and the approximation error. The two main advantages of using PQF designs are that it is easier synthesize circuits for various different universal gate sets and that the finiteness of the designs facilitates the layout of the circuit onto a fault-tolerant quantum computer architecture.

\section{Cost Analysis of PQF Circuits} \label{sec:cost:snaysis}

The optimal $T$-count has been proven to be an invariant of the unitary operation represented by a Clifford+$T$ circuit \cite{BS12,MatsumotoAmano2008,RandomRemarks}.
In particular, the optimal $T$-count is the same across various definitions of canonical and normal forms for Clifford+$T$ circuits.
At present, similar invariants have not been shown for the  Clifford+$\pi/12$ basis.
For the analysis that follows, the upper bounds proven in Appendix \ref{app:appox:pi12} suffice.

Consider a measurement of the ancilla qubits in the PQF design, such that one measurement outcome is labeled ``favorable'' and all other measurement outcomes are labeled ``unfavorable''.
Let the probability of the ``favorable''  outcome be $p$ and the unitary applied to the target qubits upon favorable measurement be $V$.
Let $C(U)$ be the cost of a circuit that performs $U$.
Assuming that the cost of performing Clifford gates is negligible, the expected $T$-count of an RUS circuit is approximately $E[C(V)] = C(U)/p$ \cite{BSRRUS}.

We assume that all rounds of the PQF circuit shown in Figure \ref{fig:pqcf-circuit} have the same probability $p_k$ of the favorable outcome that are all roughly equal to the same value, say $p$. Furthermore, we assume that the probability $q$ of unfavorable outcome satisfies $q=1-p \ll p$ and that each round roughly has the same execution cost $C_P(\varepsilon)$. We note that all these assumptions are justified by the properties of the PQF protocol derived in the following sections and obtain the following:
\begin{lem}\label{lem:cost}
For a fixed $\varepsilon$,
\begin{enumerate}
\item The expectation of the cost of the PQF protocol with $k>0$ rounds is
$$C_P(\varepsilon)/p + O(q^k). $$
\item The variance is given by
$$C_P(\varepsilon)^2   q/p^2 + O(q^k). $$
\end{enumerate}
\end{lem}

\begin{proof}
1. Let $E_k$ be the expected cost of the $k$-round protocol.
Assuming $q = 1-p << p$ we note that $p=1/p-q (p+1)/p = 1/p+O(q)$.
Clearly, we have that
\begin{eqnarray*}
E_1 &=& p   C_P(\varepsilon) + q   (C_P(\varepsilon)+C_F(\varepsilon)) \\
&=& C_P(\varepsilon)/p + O(q) C_P(\varepsilon) + q   (C_P(\varepsilon)+C_F(\varepsilon)) \\
&=& C_P(\varepsilon)/p + O(q).
\end{eqnarray*}
This provides a basis for induction on $k$.
We have $E_{k+1}= p   C_P(\varepsilon) + q   (C_P(\varepsilon)+E_k) = C_P(\varepsilon) + q   E_k$ .
By the induction hypothesis this is equal to $C_P(\varepsilon) + q   (C_P(\varepsilon)/p+O(q^k)))=(p+q) C_P(\varepsilon)/p + O(q^{k+1})$ which proves claim (1).

\noindent
2. Let $E^{(2)}_k$ be the expectation of the square of the cost. We are going to prove by induction that $E^{(2)}_k=(1+q)/p^2  C_P(\varepsilon)^2 + O(q^k) $.
For the basis of the induction we observe that
\begin{eqnarray*}
p&=&(1+q)/p^2 + p -(1+q) + O(q^2) \\
&=& (1+q)/p^2 - 2   q + O(q^2) \\
&=& (1+q)/p^2 + O(q).
\end{eqnarray*}
Therefore, we obtain that 
\begin{eqnarray*}
E^{(2)}_1 &=& p   C_P(\varepsilon)^2 + q  (C_P(\varepsilon) + C_F(\varepsilon))^2 \\
&=& (1+q)/p^2   C_P(\varepsilon)^2 + O(q)   C_P(\varepsilon)^2
\\&&+ q  (C_P(\varepsilon) + C_F(\varepsilon))^2.
\end{eqnarray*}
This in turn implies 
\begin{eqnarray*}
E^{(2)}_{k+1} &=& p  C_P(\varepsilon)^2 + q  C_P(\varepsilon)^2 + 2   q  C_P(\varepsilon) E_k + q   E(2)_k \\
&=& C_P(\varepsilon)^2 + 2   q  C_P(\varepsilon)  (C_P(\varepsilon)/p + O(q^k))\\
&&+ q   ((1+q)/p^2  C_P(\varepsilon)^2 + O(q^k)) \\
&=& ((p+q)^2 + q)/p^2  C_P(\varepsilon)^2 + O(q^{k+1}),
\end{eqnarray*}
which concludes the induction step.
Thus the variance of the cost of the $k$-round protocol is
\begin{eqnarray*}
E^{(2)}_k - E_k^2 &=&  (1+q)/p^2  C_P(\varepsilon)^2 \\
&&+ O(q^k) - (C_P(\varepsilon)/p + O(q^k))^2 \\
&=& C_P(\varepsilon)^2   q/p^2 + O(q^k).
\end{eqnarray*}
\end{proof}

\section{Overview of the PQF Algorithm}

In this section, we provide an overview of the stages of our PQF algorithm.
The algorithm returns a probabilistic quantum circuit with fallback over the chosen basis that approximates a given rotation by angle $\theta$ about the $z$-axis, denoted as $V=R_z(\theta)$, to precision $\varepsilon$.
For a multi-round PQF protocol with $k$ rounds, the algorithm sequentially generates the subcircuit for each round.
Recall that each subsequent round of the protocol is conditional on the failure of all previous rounds, and aims to ``correct" the cumulative undesired $z$-rotations and also apply the target $z$-rotation.

We develop our PQF decomposition algorithm axial $z$-rotations which then in return will allow to implement arbitrary, non-axial rotations $V$ via the Euler angle decomposition of Eq~(\ref{eq:Euler}). 

We note however that as a matter of principle, a PQF decomposition for a single-qubit unitary $V$ might be synthesizable directly without breaking $V$ into axial rotations. However, taking practical advantage of such synthesis is currently an open problem.

Our algorithm takes a predefined number $k$ of PQF rounds as input, generates primary circuit for each of the request $k$ rounds and a unitary fallback circuit that terminated the PQF protocol.

The value of $k$ can be optimally adjusted at compile time. 
Indeed, it follows from the analysis in Section~\ref{sec:cost:snaysis} that the mean expected improvement of a $k+1$-round PQF circuit on a $k$-round circuit (measured in gate count) scales down as $O((1-p)^k)$, where $p$ is the typical single round success probability. In the context of this paper we are showing that the probability can be boosted to the $\Omega(1-1/\log(1/\varepsilon))$ level (consequently in our numerical experiments even the improvement due to the second round has been insignificant).



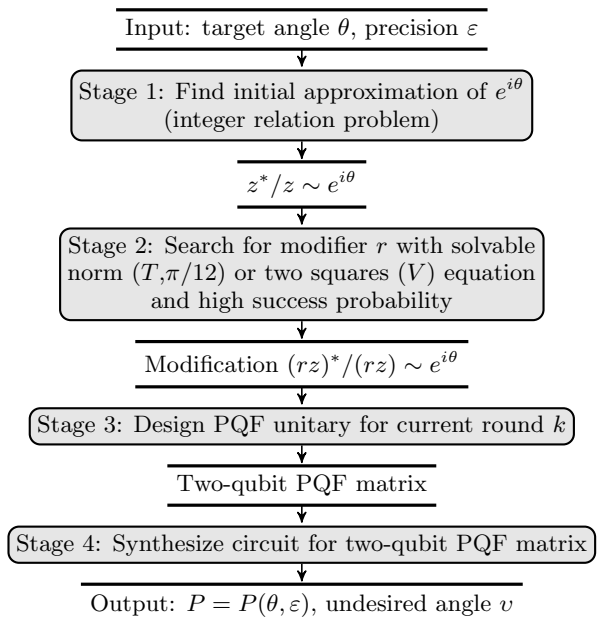
\begin{figure}[tb]
  \centering
  \makeatletter
\pgfdeclareshape{datastore}{
  \inheritsavedanchors[from=rectangle]
  \inheritanchorborder[from=rectangle]
  \inheritanchor[from=rectangle]{center}
  \inheritanchor[from=rectangle]{base}
  \inheritanchor[from=rectangle]{north}
  \inheritanchor[from=rectangle]{north east}
  \inheritanchor[from=rectangle]{east}
  \inheritanchor[from=rectangle]{south east}
  \inheritanchor[from=rectangle]{south}
  \inheritanchor[from=rectangle]{south west}
  \inheritanchor[from=rectangle]{west}
  \inheritanchor[from=rectangle]{north west}
  \backgroundpath{
    \southwest \pgf@xa=\pgf@x \pgf@ya=\pgf@y
    \northeast \pgf@xb=\pgf@x \pgf@yb=\pgf@y
    \pgfpathmoveto{\pgfpoint{\pgf@xa}{\pgf@ya}}
    \pgfpathlineto{\pgfpoint{\pgf@xb}{\pgf@ya}}
    \pgfpathmoveto{\pgfpoint{\pgf@xa}{\pgf@yb}}
    \pgfpathlineto{\pgfpoint{\pgf@xb}{\pgf@yb}}
 }
}
\makeatother
\usetikzlibrary{arrows}
\begin{tikzpicture}[
  every matrix/.style={ampersand replacement=\&,column sep=1cm,row sep=0.25cm},
  sink/.style={draw,thick,rounded corners,fill=gray!20,inner sep=.1cm},
  datastore/.style={draw,very thick,shape=datastore,inner sep=.1cm},
  to/.style={->,>=stealth',shorten >=1pt,semithick},
  every node/.style={align=center}]
 \nix{
 \matrix{
    \& \node[datastore] (ang) {Input: target angle $\theta$, precision $\varepsilon$}; \& \\
    \& \node[sink] (appr) {Stage 1: Find initial approximation of $e^{i \theta}$\\
                            (integer relation problem)}; \& \\
    \& \node[datastore] (inia) {
$z^*/z \sim e^{i \theta}$}; \& \\
 	\& \node[sink] (rsea) {Stage 2: Search for modifier $r$ with\\
                            solvable norm ($T$,$\pi/12$) or two squares ($V$) equation\\
                            and high success probability}; \& \\
	\& \node[datastore] (moda) {Modification $(r z)^*/(r z) \sim e^{i \theta}$}; \& \\
    \& \node[sink] (design) {Stage 3: Design PQF unitary for current round $k$}; \& \\
	\& \node[datastore] (matrix) {Two-qubit PQF matrix}; \& \\
 	\& \node[sink] (synt) {Stage 4: Synthesize circuit for two-qubit PQF matrix}; \& \\
	\& \node[datastore] (round) {$P=P(\theta,\varepsilon)$, undesired angle $\upsilon$}; \& \\
    \& \node[sink] (recur) {Recurse for $\{\theta - \upsilon$, $\varepsilon$, $k-1\}$ }; \& \\
    \& \node[datastore] (pop) {$F_{k-1}=PQF(\theta - \upsilon,\varepsilon, k-1)$}; \& \\
    \& \node[sink] (out) {Return $BC(failure(P),F_{k-1})$ }; \& \\
  };
}
  \matrix{
    \node[datastore] (ang) {Input: target angle $\theta$, precision $\varepsilon$}; \& \\
    \node[sink] (appr) {Stage 1: Find initial approximation of $e^{i \theta}$\\
                            (integer relation problem)}; \& \\
    \node[datastore] (inia) {
$z^*/z \sim e^{i \theta}$}; \& \\
 	\node[sink] (rsea) {Stage 2: Search for modifier $r$ with solvable\\
                            norm ($T$,$\pi/12$) or two squares ($V$) equation\\
                            and high success probability}; \& \\
	\node[datastore] (moda) {Modification $(r z)^*/(r z) \sim e^{i \theta}$}; \& \\
    \node[sink] (design) {Stage 3: Design PQF unitary for current round $k$}; \& \\
	\node[datastore] (matrix) {Two-qubit PQF matrix}; \& \\
 	\node[sink] (synt) {Stage 4: Synthesize circuit for two-qubit PQF matrix}; \& \\
	\node[datastore] (round) {Output: $P=P(\theta,\varepsilon)$, undesired angle $\upsilon$}; \& \\
  };
	\draw[to] (ang) --(appr);
	\draw[to] (appr) --(inia);
	\draw[to] (inia) --(rsea);
	\draw[to] (rsea) --(moda);
    \draw[to] (moda) --(design);
	\draw[to] (design) --(matrix);
	\draw[to] (matrix) --(synt);
	\draw[to] (synt) --(round);
\end{tikzpicture}
 \caption[Compilation algorithm]{Overview of compilation flow for one round of the PQF circuit. If a number of rounds $k$ strictly larger than one is intended, Stages 1--4 need to be repeated for modified target angles as described in the text.
}
\label{fig:intro:algorithm:T}
\end{figure}

The compilation stages for each round are outlined in Figure \ref{fig:intro:algorithm:T}. In Stage 1, detailed in Section \ref{sec:cyclotomic:approximation}, an initial approximation of the target rotation phase factor $e^{i \theta}$ is obtained.
Namely, we find an an algebraic number of the form $z^* / z$, where $z$ belongs to a set based on the chosen gate basis, to approximate $e^{i\theta}$ by finding an approximate solution to an integer relation problem.
We note that $z$ is defined up to an arbitrary real-valued factor.
The approximation is modified if needed in Stage 2, by seeking either a solvable norm equation in the case of Clifford+$T$ and Clifford+$\pi/12$ or a solvable two squares equation in the case of Clifford+$V$, and high success probability.
Stage 2 is described in Section \ref{sec:meta:modifier}.
A two-qubit unitary corresponding to the (modified) rational is composed in Stage 3 and finally synthesized into a PQF subcircuit over the chosen basis in Stage 4.
If an undesired measurement outcome occurs in the current round, the undesired rotation angle $\upsilon$ and the next target angle $\theta-\upsilon$ are generated and the latter is then used, recursively, to generate the next round of the PQF protocol.
The PQF algorithm over Clifford+$T$ and Clifford+$\pi/12$ is detailed in Sections \ref{sec:pqcf:T:and:T12}, and in Sections \ref{sec:pqcf:V} for the case of Clifford+$V$.


\section{Stage 1: Cyclotomic Rational Approximation}
\label{sec:cyclotomic:approximation}

In this section, we review the most general stage of our synthesis method.
It requires very few modifications when considering different basis sets.

Let $\zeta = e^{2  \pi   i/m}$ be the $m$-th primitive root of unity and consider the corresponding ring of cyclotomic integers $\mathbb{Z}[\zeta]$.
It is well known (c.f., \cite{LWashington}) that the minimal polynomial of $\zeta$ over rationals is monic and has degree $d = \phi(m) < m$ where $\phi$ is the Euler totient function.
We analyze the representation of an arbitrary phase factor by a unimodal cyclotomic rational $z^*/z$, where $z \in \mathbb{Z}[\zeta]$.

Let $\theta$ be a real angle.
By direct complex expansion $|z^*/z - e^{i   \theta}|= 2| \operatorname{Im} (z  e^{i   \theta/2})|/|z|$.
The phase factor $e^{i   \theta}$ is representable exactly as $z^*/z$ if and only if $\operatorname{Im} (z  e^{i   \theta/2})=0$.
It is approximately representable at precision $\varepsilon$ if and only if $|2  \operatorname{Im}(z  e^{i   \theta/2})| < \varepsilon  |z|$.
Now consider the standard integer basis $\{1,\zeta,\ldots,\zeta^{d-1}\}$ in $\mathbb{Z}[\zeta]$.
Representing $z$ in this basis results in
$z=a_0+a_1  \zeta + \ldots + a_{d-1}  \zeta^{d-1}$, where $\{a_0, a_1,  \ldots, a_{d-1}\}$ are ordinary integers.

Again, by direct complex expansion we observe that $\operatorname{Im}(z   e^{i   \theta/2})$ is a linear form with real coefficients in
$\{a_0, a_1,  \ldots, a_{d-1}\}$.
We expand this form as
$F(a,x(\theta)) = a_0   x_0(\theta) +a_1  x_1(\theta) + \ldots + a_{d-1}  x_{d-1}(\theta)$, where
$x_j(\theta)=\sin(\theta/2+2  \pi   j/m)$ and $j=0,\ldots, d-1$ is the corresponding real vector.
It is easy to see that for $\theta$ in a general position, the vector does not have zero components.
It is also helpful to observe that for $|\theta|< \pi/2$ at least one $x_j$ is well separated from zero (e.g., at least one $x_j(\theta)$ has to be greater than $\sin(2   \pi/m)$).

Representing the phase factor $e^{i   \theta}$ exactly as a cyclotomic rational is equivalent to solving an integer relation with real coefficients, namely solving $F(a,x(\theta)) = 0$ for $a$.
Furthermore, when it is not solvable we consider finding approximate integer relations, i.e., finding $\{a_0, a_1,  \ldots, a_{d-1}\}$ that $|F(a,x(\theta))| < \delta$.
It is well known \cite{FergBail} that such approximate relations can be algorithmically found for arbitrarily small positive $\delta$.

\begin{lem} \label{interger:relation:lemma}
For a fixed $\theta$ in a general position, $|\theta|< \pi/2,$ and sufficiently small $\delta>0$, there exists an integer solution $a$ of $|F(a,x(\theta))| < \delta$ such that $|a_j|=O(\delta^{-1/(d-1)})$, $j=0,\ldots, d-1$.
\end{lem}

\begin{proof}
The proof follows from a more general theorem regarding the quality of multivariate Diophantine approximations (c.f.,~\cite{WSchmidt}, Section II, Theorem 1C):
For any real numbers $x_1,...,x_n$ and $0 < \epsilon < 1$ there exist integers $q_1, \ldots, q_n, p$ such that $|q_1   x_1 + \cdots + q_n   x_n - p | < \epsilon$ and $\max(|q_1|, \ldots, |q_n|)< \epsilon^{-1/n}$.

We apply this theorem to our case for $n=d-1$.
As observed, at least one of the coefficients $x_j(\theta)=\sin(\theta/2 + 2  \pi   j/m)$ is in the interval $(\sin(2 \pi/m), 1)$.
We can relabel the $x_j(\theta)$ for convenience so that one of the coefficients belonging to the interval $(\sin(2 \pi/m), 1)$ is labeled $x_0(\theta)$.
Set $x_j=x_j(\theta) / x_0(\theta)$, $j = 1, \ldots d-1$.
%
By applying Theorem 1C, we conclude that there exists an integer solution $a$ of $|F(a,x(\theta))|/|x_0(\theta)| < \epsilon $ with $|a_j| < \epsilon^{-1/(d-1)}$, $j = 1, \ldots, d-1$.

By the triangle inequality, $|a_0| \leq |a_1|   |x_1|+\ldots + |a_{d-1}|   |x_{d-1}| + \epsilon$, where $\epsilon$ is negligibly small compared to $\epsilon^{-1/(d-1)}$ and where, by design, $|x_j| < 1/|x_0(\theta)|$, $j=1, \ldots, d-1$.
Thus $|a_0| < (d-1)/|x_0(\theta)|  \epsilon^{-1/(d-1)}$.
Setting $\epsilon$ to be smaller than $\delta / |x_0(\theta)|$ concludes the proof of the lemma.

\end{proof}

\begin{corol} \label{corol:integer:rel:general}
For a fixed $\theta$ in a general position, $|\theta|< \pi/2$, and sufficiently small $\varepsilon>0$, there exists a cyclotomic rational approximation $|z^*/z - e^{i\theta}| < \varepsilon$, $z \in \mathbb{Z}[\zeta]$ with $|z|$ in $O(\varepsilon^{-1/d})$.
\end{corol}

\begin{proof}
Per Lemma \ref{interger:relation:lemma}, a solution $z$ to $|z^*/z - e^{i\theta}| < \varepsilon$ must exist.
Setting $\delta = \varepsilon   |z|/2$ in the lemma, we infer the existence of a solution $z$ in $O(|z|^{-1/(d-1)}   \varepsilon^{-1/(d-1)})$.
This implies $|z|^{d/(d-1)}$ in $O( \varepsilon^{-1/(d-1)})$, or $|z|^d$ in $O(\varepsilon^{-1})$, and the corollary follows.
\end{proof}

In order to find the solutions algorithmically, we customize the PSLQ integer relation algorithm \cite{FergBail,PSLQBertok}.
PSLQ is an iterative algorithm to solve integer relations of the form $a.x = a_1  x_1 + \ldots + a_d  x_d$, where $|a.x|$ can be made arbitrarily small after a large enough number of iterations.
Our customization terminates when the equivalent of the $|z^*/z-e^{i \theta}| < \varepsilon$ inequality is first satisfied.
The performance proofs in \cite{FergBail,PSLQBertok} can be modified to show that $|z|$ upon termination is in $O(\varepsilon^{-1/d})$.

Our numerical experiments provide an estimate of the asymptotics of $|z|$ in cases $m=4,8,12$ (which correspond to the Clifford+$V$, Clifford+$T$, and Clifford+$\pi/12$ bases, respectively).
For example, when $m=8$ (Clifford+$T$) we find $|z| < \kappa \varepsilon^{-1/4}$, where $\kappa = 3.05 \pm 0.28$.

The following observation will be necessary for compilation Stage 2 in designing a matrix over the Clifford+$T$ and Clifford+$\pi/12$ bases. In that context not only the size of $z$ but also the size of its ``Galois conjugate'' $z^{\bullet}$ comes into play.

\begin{observ} \label{the:bullet:observ}

In the context of Corollary \ref{corol:integer:rel:general}, consider $z^{\bullet}$ which is obtained from $z$ by formally replacing $\omega$ by $-\omega$. That is, $z^{\bullet} = a_0+a_1(-\omega)+\ldots +  a_{d-1}   (-\omega)^{d-1}$ when $z = a_0+a_1 \omega+\ldots +   a_{d-1}   \omega^{d-1}$.
The $z^{\bullet}$ is also in $O(\varepsilon^{-1/d})$.

\end{observ}

Indeed the proof of Lemma \ref{interger:relation:lemma} relies only on the bounds for absolute values of the coefficients, and does not change if only the signs of the coefficients are altered.
Thus $z^{\bullet}$ is in $O(|z|^{-1/(d-1)}   \varepsilon^{-1/(d-1)})$ and the observation follows.

\begin{observ} \label{observe:size:lower:T}
In the context we can assume, without loss of generality that $\log(|z|) > 1/(2\,d) \,  \log(1/\varepsilon)$.
\end{observ}

If this is not the case, pick the integer $s = \lceil \varepsilon^{-1/(2\,d)} / |z| \rceil$ and replace $z$ with $s\,z$.

\section{Stage 2: Search for a Modifier} \label{sec:meta:modifier}

In this section, we present the design of our algorithm and supporting mathematical rigor required for applying the algorithm to each gate basis.

Following the previous section, let $\zeta = e^{2  \pi   i/m}$.
We limit our analysis to values of $m$ that are multiples of $4$ such that the ring $\mathbb{Z}[\zeta]$ contains $i = \sqrt{-1}$.

We introduce the \emph{unitarization base} $\nu$, where $\nu = \sqrt{2}$ for $m>4$ and $\nu=\sqrt{5}$ for $m=4$ (the latter will be relevant for the $V$-basis).
Let $\theta$ be the target angle of rotation about the $Z$-axis and $z^*/z$, where $z \in \mathbb{Z}[\zeta]$, be an $\varepsilon$-approximation of the phase factor $e^{i  \theta}$.

The synthesis of both purely unitary and measurement-assisted decomposition circuits
hinge on the existence of a unitary matrix of the form
\begin{equation} \label{general:z:y:matrix}
\frac{1}{\nu^L}   \left[\begin{array}{cc}
              z & y \\
              -y^* & z^*
            \end{array}\right],
\end{equation}
where $y \in \mathbb{Z}[\zeta]$, and $L \in \mathbb{Z}$.
The unitary condition for this matrix, $|y|^2 = \nu^{2 L} - |z|^2$, is restrictive; the matrix need not exist for an arbitrary $z$ as $z$ being part of a cyclotomic rational approximation does not imply its existence.

For measurement-assisted circuit decomposition, another constraint is also relevant.
Assuming the unitary matrix of the form (\ref{general:z:y:matrix}) exists, we may introduce $p_1 = |z|^2/\nu^{2 L} < 1$.
For the measurement-assisted circuits to have sufficient quality we will need $p_1$ to be greater than $1-O(1/L)$, which we show below.

For $m \neq 4$, we introduce $\rho = \zeta + \zeta^*$ and the real subring $R= \mathbb{Z}[\rho] \subset \mathbb{Z}[\zeta]$.
In the special case of $m=4$, we set $R=\mathbb{Z}$ (note that $\rho=0$ in this special case and we would be reluctant to argue that $\mathbb{Z}[0]=\mathbb{Z}$.).
Any element $r \in R$ evaluates to a real number.

To address both of the above constraints in one design, we note that for any non-zero $r \in R$, we have $(r   z)^*/(r   z) = z^*/z$.
Thus replacing $z$ with $r z$ does not change the cyclotomic approximation.

\begin{lem} \label{prob:modifier:lemma} (Meta-statement)
In the above context, let $L_1 = \lceil \log_{\nu}(|z|) \rceil$.
There exists an algorithmically defined subset $S_{z} \subset R$ of cardinality $\Theta(L_1)$ such that for any $r \in S_z$
\begin{enumerate}
\item $0 < \lceil \log_{\nu} (|r z|) \rceil - \log_{\nu} (| r z|) < O(1/L_1)$, and
\item $\log_{\nu}(|r|)$ is in $O(\log(L_1))$.
\end{enumerate}
\end{lem}
We currently do not have a proof of this lemma for arbitrary cyclotomic ring.

Assuming Lemma \ref{prob:modifier:lemma} holds, for any $r \in S_z$  let
$L_r = \lceil \log_{\nu} (|r z|) \rceil$.
Claim (1) of Lemma \ref{prob:modifier:lemma} directly implies that $p_r = |r   z|^2/ \nu^{2   L_r} > 1 - O(1/L_1)$ while claim (2) implies that $L_r$ is in $L_1+O(\log(L_1))$.
There is also an algorithmically defined set of at least $\Theta(L_1)$ values with these properties.

Intuitively, in the subsequent designs for the probabilistic measurement-assisted circuits $p_r > 1 - O(1/L_1)$ means there will be a high one-round success rate.
The asymptotics for $L_r$ implies that no $r$ from $S_{z}$ will substantially increase the depth of the resulting circuit.

The existence of the matrix
\begin{equation} \label{general:r:z:y:matrix}
\frac{1}{\nu^{L_r}}   \left[\begin{array}{cc}
              r z & y \\
              -y^* & r z^*
            \end{array}\right],
\end{equation}
for some chosen  $r \in S_z$, is equivalent to solving the equation
\begin{equation} \label{deja:norm:eq}
|y|^2 = \nu^{2 L_r} - | r   z|^2,
\end{equation}
for $y \in \mathbb{Z}[\zeta]$.
Eq (\ref{deja:norm:eq}) is known as a \emph{norm equation} over the cyclotomic integers. Its solvability and solutions are well understood \cite{LWashington}.

Consider the \emph{absolute norm} map $N: \mathbb{Q}(\rho) \rightarrow \mathbb{Q}$. It is a general fact that $N(R) \subset \mathbb{Z}$.
We use terminology and facts from  \cite{LWashington} in our below description.
We first address a particular case where $p=N(\nu^{2 L_r} - | r   z|^2)$ is a prime integer.
As per Theorem 2.13 \cite{LWashington}, the norm equation in Eq (\ref{deja:norm:eq}) is solvable if and only if  $p = 1  \mod  m$.
Intuitively this means that solvable norm equations are not rare.
Let $B$ be an arbitrarily large positive integer.
It is well known that the density of prime numbers in, say, the segment $[B/2,B]$ is in $\Omega(1/ln(B))$.
It is also well know that if $m << B$ then the density of such prime numbers $p$ such that $p=1   \mod   m$ in that segment is still in $\Omega(1/ln(B))$.

Suppose we have identified a large enough subset $S_z \subset R$ so that the set of integers
$\{N(\nu^{2 L_r} - | r   z|^2) | r \in S_z\}$
intersects with some segment of the form $[B/2, B]$ and the intersection has $\Theta(ln(B))$ distinct integers, i.e., the number of distinct integers in the intersection is $ln(B)$ times some significant factor.
Then with some high probability there is an $r \in S_z$ such that $p=N(\nu^{2 L_r} - | r   z|^2)$  is prime and $p = 1   \mod  m$.

However, this is a minimalistic approach.
If $p=N(\nu^{2 L_r} - | r   z|^2)$ is not prime, and its prime factorization is known, then the complete analysis of solvability of Eq (\ref{deja:norm:eq}) can be algorithmically performed in polynomial time.
Therefore we can broaden the search for feasible values of $r$ by looking at such values where $N(\nu^{2 L_r} - | r   z|^2)$ is easy to factor (e.g., it is a smooth integer).
We refer to Eq (\ref{deja:norm:eq}) as \emph{easily solvable} when $N(\nu^{2 L_r} - | r   z|^2)$ is easy to factor \emph{and} the equation has a solution.

\begin{cnj} \label{meta:conj} (Meta-conjecture)

For any $z \in \mathbb{Z}[\zeta]$ in a general position there exists a certain subset $S_z \subset R$ that satisfies the claims of Lemma \ref{prob:modifier:lemma}, has cardinality in $O(L_1)$, and contains at least one $r$ for which the equation (\ref{deja:norm:eq}) is easily solvable.

\end{cnj}

Assuming this conjecture holds, we can manufacture a unitary matrix of the form (\ref{general:r:z:y:matrix}) in polynomial classical runtime with the promise that its $L_r$ is in $\log_{\nu}(|z|) + O(\log(\log(|z|))$.
We proceed by describing how to apply the described framework in the context of each of the three basis sets.

\section{PQF over Clifford+$V$} \label{sec:pqcf:V}

Unitary decomposition of single-qubit rotations over the so-called $V$-basis was described in Ref.~\cite{GENUINE}.
While RUS decomposition over Clifford+$V$ has not yet been shown, the PQF protocol allows generalization to the $V$ basis.
We generalize PQF to Clifford+$V$ in this section, and show that it is remarkably simpler than in the Clifford+$T$ case.

Recall that the single-qubit $V$ gate is given by
$V=(I-2   i   Z)/\sqrt{5}$.
The group of circuits generated by the Clifford group and the $V$ gate is universal for quantum computation (ibid.).
It has also been shown that an arbitrary single-qubit unitary gate can be approximated to precision $\varepsilon$ by a single-qubit unitary Clifford+$V$ circuit with $V$-count bounded by $3 \,  \log_5(1/\varepsilon) + O(\log(\log(1/\varepsilon)))$ \cite{BGV}.
For approximation of axial rotations, the algorithm is efficient.
For arbitrary single-qubit targets, the same $V$-count can be achieved using an exponential-time algorithm (ibid.).
that is practically feasible for a reasonable range of precisions.

The guarantees for the unitary decomposition algorithm are based on the following fact, which is the basis for our efficient algorithm to approximate an arbitrary axial rotation $R_z(\theta)$ with a PQF circuit over Clifford+$V$:
\begin{lem} \label{v:exact:decomposition:lemma}
A unitary matrix of the form
$\frac{1}{{\sqrt{5}}^{L}}   \left[\begin{array}{cc}
              z & y \\
              -y^* & z^*
            \end{array}\right]$,
where $y,z$ are Gaussian integers and $L \in \mathbb{Z}$, can be exactly and algorithmically decomposed into a Clifford+$V$ circuit of $V$-count at most $L$.
\end{lem}


Following Lemma \ref{v:exact:decomposition:lemma}, we consider the case $\zeta=i$ (which corresponds to $m=4$). Here $\mathbb{Z}[i]$ is a quadratic extension of $\mathbb{Z}$ so $d=2$.
By Corollary \ref{corol:integer:rel:general}, any phase factor $e^{i   \theta}$ can be approximated with a Gaussian rational $z^*/z, z \in \mathbb{Z}[i]$, where $|z|$ is in $O(\varepsilon^{-1/2})$.

\begin{observ} \label{observe:lower:size:V}

Without loss of generality we can assume that $\log(|z|) > 1/4   \log(1/\varepsilon)$.

\end{observ}
If it is not the case, we pick the integer $s = \lceil \varepsilon^{-1/4} /|z| \rceil$ and replace $z$ with $s  \, z$.

\subsection{Stage 2: Probability Modifier}

We now prove Lemma \ref{prob:modifier:lemma} over Clifford+$V$.
%
Let $L_1= \lceil \log_{\sqrt{5}}(|z|) \rceil$.
We want to define a subset $S_z \subset \mathbb{Z}$ such that
$\forall r \in S_z, \lceil \log_{\sqrt{5}}(|r   z|) \rceil - \log_{\sqrt{5}}(|r   z|) < 1/L_1$.
Let $\lambda = L_1 - \log_{\sqrt{5}}(|z|)$.
We select values of $r$ such that $0< \log_{\sqrt{5}}(|r|) - \lfloor \log_{\sqrt{5}}(|r|) \rfloor < \lambda$.
Under this assumption $\lceil \log_{\sqrt{5}}(|r   z|) \rceil - \log_{\sqrt{5}}(|r   z|) = \lambda - (\log_{\sqrt{5}}(|r|) - \lfloor \log_{\sqrt{5}}(|r|))$.

We define the desired $S_z$ as a subset of positive integers $r$ satisfying the inequality
\begin{equation*}
\lambda - 1/L_1 < \log_{\sqrt{5}}(|r|) - \lfloor \log_{\sqrt{5}}(|r|) < \lambda.
\end{equation*}
It is necessary and sufficient that  $\log_{\sqrt{5}}(|r|)$ is in an interval of the form $(k + \lambda - 1/L_1, k + \lambda)$, where $k \in \mathbb{Z}$ or $|r|$ is in the interval
We have $I_k=(\sqrt{5}^k   \sqrt{5}^{\lambda - 1/L_1}, \sqrt{5}^k   \sqrt{5}^{\lambda})$.
It follows that the number of integers contained in $I_k$ grows exponentially with $k \geq  k_0 = \lceil \log_{\sqrt{5}}(L_1) \rceil$.

We define the desired $S_z$ as the ordered sequence of all integers in $\Cup_{k\geq k_0} I_k$.
While $S_z$ is an infinite sequence, any of its initial subsequences of length $O(L_1)$ is contained in $\Cup_{k_0 \leq  k \leq O(\log(L_1))} I_k$.
This concludes the proof.


Next we specialize Conjecture \ref{meta:conj} for $m=4$ and discuss its implications.
Let $S_z'$ be some initial subsequence of length $\Theta(L_1)$ in $S_z$ and let $S_z'(easy)\subset S_z'$ be the subsequence of such $r \in S_z'$ for which the norm equation $|y|^2 = 5^L_r - |r z|^2$ is easily solvable for $y \in \mathbb{Z}[i]$.

\begin{cnj}\label{cnj:V}
The density of $S_z'(easy)$ in $S_z'$ is in $\Omega(1/\log(L_1))$.
It suffices to inspect $O(L_1)$ initial values in $S_z$ in order to find one for which the norm equation is easily solvable.
\end{cnj}

Conjecture \ref{cnj:V} implies that we need to test at most $O(L_1)=O(\log(|z|))=O(\log(1/\varepsilon))$ norm equations for easy solvability to find one that is easily solvable.
It also implies that the value of $L_r$ corresponding to the solution is in
$L_1 + O(\log(L_1)) = \log_{\sqrt{5}}(|z|) + O(\log(\log(|z|))) = \log_5(1/\varepsilon)+ O(\log(\log(1/\varepsilon)))$

\subsection{Stages 3, 4: Design and Synthesis of PQF Subircuits}

We now have the unitary matrix
$W = \frac{1}{{\sqrt{5}}^{L_r}}   \left[\begin{array}{cc}
              r z & y \\
              -y^* & r z^*
            \end{array}\right],$
where $L_r = \lceil \log_5(r^2 |z|^2) \rceil \leq \log_5(|z|^2) + O(\log(\log(|z|^2)))$ and $r^2 |z|^2 / 5^{L_r} > 1 - 1/\log_5(|z|^2).$

As observed in \cite{GENUINE}, $V$ is exactly represented by a unitary Clifford+$V$ circuit with $V$-count at most $L_r$.
Therefore, the two-qubit PQF matrix $U = \mbox{CNOT}   (I \otimes W)  \mbox{CNOT}$ is exactly represented by a circuit with the same $V$-count.

By direct computation, when $U$ is applied to $|\psi \rangle |0\rangle$ and the second qubit is measured then either
\begin{itemize}
\item on measurement outcome $0$ the $\Lambda(z^*/z) \sim \Lambda(e^{i   \theta})$ rotation gate is effectively applied to the primary qubit, or
\item on measurement outcome $1$ the $\Lambda(-y/y^*)$ rotation gate is applied to the primary qubit.
\end{itemize}
Thus the round one fallback circuit must be a unitary $\varepsilon$-approximation of the rotation gate $\Lambda(-y^*/y   e^{i   \theta})$.
Fallback circuits at subsequent rounds have similar structure.

As per \cite{BGV} any of the fallback circuits can be implemented at $V$-count of at most $3  \log_5(1/\varepsilon) + O(\log(\log(1/\varepsilon)))$.
Following the guarantees derived for Stage 1 of this algorithm, $|z|^2$ is in $O(1/\varepsilon)$ therefore the $V$-count of the two-cubit circuit for $U = \mbox{CNOT}   (I \otimes V)   \mbox{CNOT}$ is bounded by $\log_5(1/\varepsilon) + O(\log(\log(1/\varepsilon)))$.

The one round ``failure" rate of the circuit, which is the probability $q$ of measuring $1$, is less than $1/\log_5(|z|^2)$.
As per Observation \ref{observe:lower:size:V}, we can assume that $\log_5(|z|^2) > 1/2   \log_5(1/\varepsilon)$ and therefore $q < 2/\log_5(1/\varepsilon)$.
Thus the expected $V$-count of the one-round PQF protocol is bounded by $\log_5(1/\varepsilon) + O(\log(\log(1/\varepsilon))) + q \,  (3  \log_5(1/\varepsilon) + O(\log(\log(1/\varepsilon)))) < \log_5(1/\varepsilon) + O(\log(\log(1/\varepsilon)))$.
Similarly, the $V$-count of the two-round PQF protocol is bounded by $\log_5(1/\varepsilon) + O(\log(\log(1/\varepsilon)))$.

\section{PQF over Clifford+$T$ and Clifford+$\pi/12$} \label{sec:pqcf:T:and:T12}

The Clifford+$T$ basis is arguably the most popular universal quantum basis \cite{IkeAndMike2000}.
It consists of the multi-qubit Clifford group and the single-qubit  $T$ gate, where $T = \left[\begin{smallmatrix}1&0\\0&e^{i \pi/4}\end{smallmatrix}\right]$.
Alternatively, the basis can be viewed as being generated by $\{T, H, \mbox{CNOT}\}$.
We cost our synthesized circuits by the number of $T$ gates, which is motivated by the high cost of fault-tolerant implementations of the $T$ gate (or other non-Clifford gate) \cite{Bravyi2004,MeierEtAl,BravyiHaah}.

The Clifford+$\pi/12$ basis analogously consists of the multi-qubit Clifford group and the single-qubit  $K=\pi/12$ gate, where $K = \left[\begin{smallmatrix}1&0\\0&e^{i \pi/6}\end{smallmatrix}\right]$.
It is generated by the set $\{K, H, \mbox{CNOT}\}$.
The study of this set is motivated by recent results on the universality of metaplectic anyons \cite{Metaplectic}.
While we present an algorithm to decompose into this basis, an efficient fault-tolerant implementation of the $K=\pi/12$ gate, for example by magic state distillation, remains open for future research.
In this analysis, however, we assume, paralleling the Clifford+$T$ basis, that the cost of executing a $\pi/12$ gate is significantly higher than the cost of executing a Clifford gate.
We also assume that $K$, $K^{-1}$, $K^2$, and $K^{-2}$  have the same unit cost.
Therefore the cost of a Clifford+$\pi/12$ circuit is dominated by the number of $K$-monomials occurring in the circuit.

In all other technical aspects the Clifford+$T$  and Clifford+$\pi/12$ systems are strikingly similar.
For the Clifford+$T$ and Clifford+$\pi/12$ bases, the first stage of the algorithm approximates the phase factor $e^{i \theta}$ with a unimodal cyclotomic rational, i.e., an algebraic number of the form $z^* / z$, where $z \in \mathbb{Z}[\omega]$, by finding an approximate solution of an integer relation problem.
The second stage performs the modification $z \mapsto (r z)$, where $ r \in \mathbb{Z}[\rho]$ using Lemma \ref{r:sampling:lemma} developed below.
The third and fourth stages design and synthesize the PQF subcircuit.
We review the algorithm in the case of these two bases below.

\subsection{Exactly Representable Unitaries}

We denote $\omega = e^{i \pi /4}$ and $\omega_{12} = e^{i \pi /6}$. Both $\omega$ and $\omega_{12}$ are algebraic integers of degree $4$.
We intentionally omit the superscript $12$ and use $\omega$ when the algebra of these two algebraic integers is identical, and denote it otherwise.
We denote $\rho = \omega+\omega^*= \sqrt{2}$ and $\rho_{12} = \omega_{12}+\omega_{12}^*= \sqrt{3}$.
Again we use $\rho$ without subscript when no distinction is necessary.

The fundamental unit $\upsilon = 1 + \rho$ of the $\mathbb{Z}[\rho]$ ring and the fundamental unit
$\upsilon_{12} = 2 + \rho_{12}$ of the $\mathbb{Z}[\rho_{12}]$ ring.
We use $\upsilon$ without subscript when distinction is unnecessary.

The algebraic number ring $\mathbb{Z}[\omega]$, is a degree $4$ extension of $\mathbb{Z}$.
The Galois group of this extension is the direct product $\mathbb{Z}_2 \times \mathbb{Z}_2$ generated by complex conjugation $*$ and one other automorphism $\bullet$ that extends $\omega^{\bullet} = - \omega$.
The ring $\mathbb{Z}[\omega]$ has an \emph{integer basis} of four elements, with the most obvious basis being $\{ \omega^3 , \omega^2 , \omega, 1 \}$ \cite{KMM12}.
It consists of all numbers of the form $ a   \omega^3 + b   \omega^2 + c   \omega + d$, where $a,b,c,d$ are arbitrary integers.

It was shown in \cite{KMM12} for the Clifford+$T$ system that a unitary operation $V$ on $n$ qubits is representable exactly by a Clifford+$T$ circuit if and only if it is of the form $V = 1/\sqrt{2}^k M$, where $M$ is a matrix with elements from $\mathbb{Z}[\omega]$ and $k$ is some non-negative integer.
To satisfy the unitary condition, we require $M   M^{\dagger} = 2^k   \onemat_{2^n}$.
Moreover, it was shown that a matrix of this form can be represented as an asymptotically optimal Clifford+$T$ circuit using at most two ancilla qubits \cite{GilSel,Kliuchnikoff}, and no ancilla qubits when either the target is a single-qubit unitary or when $\det(1/\sqrt{2}^k M)=1$~\cite{GilSel}.

In Appendix \ref{app:exact:pi12} we extend this claim to single-qubit unitaries over the Clifford+$\pi/12$ basis.
We prove that a $V \in U(2)$ is representable exactly as a Clifford+$\pi/12$ circuit if it is of the form $V = 1/\sqrt{2}^k M$, where $M$ is a $2 \times 2$ matrix over $\mathbb{Z}[\omega_{12}]$ such that $M   M^{\dagger} = 2^k   \onemat_{2}$

We note that the PQF and RUS protocols share that when the phase factor $e^{ i   \theta}$ is approximated by some $y/{\sqrt{2^\ell}}$, where $y \in \mathbb{Z}[\omega]$, and by some $z^*/z$, where $z \in \mathbb{Z}[\omega]$ for the same precision, then $z$ will in general have much smaller bitsize than $y$ for that precision.

\subsection{Stage 2: Probability Modifier}
\label{sec:randomized:search}

Let $z^*/z, z \in \mathbb{Z}[\omega]$ be a cyclotomic rational approximation of $e^{i \theta}$ as explained in Section \ref{sec:cyclotomic:approximation}.
In Stage 2, we include $z$ in a unitary of the form (\ref{general:z:y:matrix}), where in this context $\nu=\sqrt{2}$, $y \in \mathbb{Z}[\omega]$, and $L \in \mathbb{Z}$.
We would like $|z|^2/2^L$ to be reasonably large since this value equals the success probability of the current round in the PQF protocol.
Unfortunately, the majority of $z$ values do not allow this.
To create a unitary of the form (\ref{general:z:y:matrix}), we seek a $y$ that satisfies the normalization condition $(|y|^2+|z|^2)/2^L = 1$, or equivalently $|y|^2 = 2^L - |z|^2$.
It is easy to see that $|z|^2$ belongs to the real-valued ring $\mathbb{Z}[\rho]$ and thus so does $2^L - |z|^2$.

Given an arbitrary $\xi \in \mathbb{Z}[\rho]$, the identity
\begin{equation}\label{eq:norm:equation}
|y|^2 = \xi,
\end{equation}
considered as an equation for an unknown $y \in \mathbb{Z}[\omega]$, is called a \emph{norm equation in} $\mathbb{Z}[\omega]$.
Deciding whether a given norm equation is solvable and finding a solution is in general at least as hard as performing factorization of an arbitrary integer.
For our algorithm to be efficient, we need to find norm equations that are easy to solve.

A necessary solvability condition to construct a matrix of the form (\ref{general:z:y:matrix}) with $\nu=\sqrt{2}$ is given by $|z|^2 \leq 2^L$ and $|z^{\bullet}|^2 \leq 2^L$, where $()^{\bullet}: \mathbb{Z}[\omega] \rightarrow \mathbb{Z}[\omega]$ extends the map $\omega \mapsto (-\omega)$.
We can arbitrarily replace $z$ in $z^*/z$ by $r  z$, where $r \in \mathbb{Z}[\rho]$, without changing the fraction.

Our strategy generalizes that of Ref.~\cite{Selinger}.
Consider a fixed $z \in \mathbb{Z}[\omega]$. Introduce $L_1 = \lceil \log_2(|z|^2) \rceil$.
Note that for $z$ defined at Stage 1, $L_1$ follows the asymptotics of $\log_2(1/\varepsilon)/2+c$, where $c$ is a constant.
Thus for asymptotically small $\varepsilon$, $L_1$ is asymptotically large.
For a randomly chosen $r \in \mathbb{Z}[\rho]$, we set $L_r = \lceil \log_2(|r   z|^2) \rceil$.

Further design and analysis of our algorithm is based on the following:
\begin{cnj} \label{cnj:enough:normeq}
In the above context, consider the set $S_{+}(z) = \{r \in \mathbb{Z}[\rho] \, | \, 2^{L_r}-|r  z|^2 \geq 0 ,\,  2^{L_r}-|(r  z)^{\bullet}|^2 \geq 0\}$.
Let $S_{solvable}(z) \subset S_{+}(z)$ be the subset of such $r \in \mathbb{Z}[\rho]$ for which the norm equation $|y|^2 =  2^{L_r}-|r  z|^2$ is solvable in $\mathbb{Z}[\omega]$.
Then the average density of $S_{solvable}(z)$ in $S_{+}(z)$ belongs to $\Omega(1/L_1)$, when $|z|\rightarrow \infty$ and $ L_1 \rightarrow \infty$.
\end{cnj}

This conjecture is a special case of Conjecture \ref{meta:conj}.
The motivation for conjectures of this type was discussed in Section \ref{sec:meta:modifier}.
In our numeric experiments for over $1000$ random angles and $30$ levels of precision we have not encountered a single failure, suggesting the conjecture did not hold.

\begin{lem} \label{r:sampling:lemma}
For sufficiently large $L_1$,  an $r \in \mathbb{Z}[\rho]$ can be algorithmically found, in a probabilistically polynomial number of steps, such that
\begin{enumerate}
\item $L_r \leq L_1 + \log_2(|z^{\bullet}/z|) + O(\log(L_1))$,
\item $p(r)> |r   z|^2/2^{L_r} > 1 - 1/L_1$,
\item $2^{L_r} - |r^{\bullet}   z^{\bullet}|^2 \geq 0$,
\item  the norm equation $|y|^2 = 2^{L_r} - |r   z|^2$ is easily solvable in $\mathbb{Z}[\omega]$.
\end{enumerate}
\end{lem}

\begin{proof}

As per Conjecture \ref{cnj:enough:normeq} it suffices to find a set of values $r$ that satisfy (1) -- (3) in Lemma \ref{r:sampling:lemma} of size $O(L_1)$.
Then there is at least one value $r$ in a set of size $O(L_1)$ for which the norm equation is easily solvable.

Before  algorithmically constructing such a set, we note that condition 3 is somewhat redundant.
It is a necessary condition for the norm equation to be solvable and conveniently helps reduce the search space for $r$.

Let $\zeta = \log_2(2^{L_1}/|z|^2)=L_1-\log_2(|z|^2)$.
We rewrite conditions (1) and (2) in terms of $R= \lfloor \log_2(|r|^2) \rfloor$ and $f = \log_2(|r|^2)-R$.
The first condition can be restated as $R$ is in $\log_2(|z^{\bullet}/z|) +O(\log(L_1))$.
The second condition means that $2^{-(\zeta-f)} > 1 - 1/L_1$.
We subsequently design $f$ to be smaller than, but close enough to, $\zeta$.
It follows that $(1 - 1/L_1) 2^{\zeta} < 2^{f} < 2^{\zeta}$.

As per the definitions of $\zeta, f, R$ a slightly stronger (asymptotically equivalent) condition on $|r|$ is given by
\begin{equation*}
(1-1/(2 L_1))  2^{(R+\zeta)/2} < |r| < 2^{(R+\zeta)/2}.
\end{equation*}
We then rewrite condition 3 as
$|r^{\bullet}|\leq 2^{L_1/2}/|z^{\bullet}| \times 2^{R/2}$.

We are now ready to describe the construction of a sufficient set of values $r$.
This is one of the few places where the distinction between Clifford+$T$ and Clifford+$\pi/12$ must be made.
Recall the fundamental unit $\upsilon=1+\rho$ we have defined for Clifford+$T$ and the fundamental unit $\upsilon_{12}=2+\rho_{12}$ defined for Clifford+$\pi/12$.

Our construction for Clifford+$T$ exploits Lemma 17 from \cite{Selinger}: given real numbers $x_0,x_1,y_0,y_1$ such that $|(x_1 - x_0) (y_1 - y_0)| > \upsilon^2$, one can algorithmically find an $r \in \mathbb{Z}[\sqrt{2}=\rho]$ such that $r \in (x_0, x_1)$ and $r^{\bullet} \in (y_0, y_1)$.

The construction for Clifford+$\pi/12$ is based on a version of the lemma developed in Appendix \ref{app:appox:rt3}: given real numbers $x_0,x_1,y_0,y_1$ such that $|(x_1-x_0) (y_1-y_0)|> \upsilon_{12}^2$, one can algorithmically find an $r \in \mathbb{Z}[\sqrt{3}=\rho_{12}]$ such that $r \in (x_0,x_1)$ and $r^{\bullet} \in (y_0,y_1)$.

The two lemmas are identical except for the $12$ subscript. We omit the subscript in the following narration which is common for the two cases. We collectively refer to the two lemmas as the ``bullet lemmas".

Set $x_0(R) = \left(1-1/(2 L_1)\right)  2^{(R+\zeta)/2}$, $x_1(R) = 2^{(R+\zeta)/2}$, $y_0(R)=-2^{L_1/2}/|z^{\bullet}| \times 2^{R/2}$, and $y_1=+2^{L_1/2}/|z^{\bullet}| \times 2^{R/2}$.
Then $|(x_1(R) - x_0(R)) (y_1(R) - y_0(R))|=2^R |z/z^{\bullet}|   1/L_1$. If the latter value is greater than $\upsilon^2$ , or equivalently,
$2^R > \upsilon^2   |z^{\bullet}/z|   L_1 \in O(L_1)$
then one can algorithmically find at least one $r \in \mathbb{Z}[\rho]$ that satisfies conditions 1 -- 3.

Consider $R_0(z) = \lceil \log_2(\upsilon^2   |z^{\bullet}/z|   L_1) \rceil$.
Obviously $R_0(z)$ is in $\log_2(|z^{\bullet}/z|   L_1)+O(\log(L_1))=\log_2(|z^{\bullet}/z|)+O(\log(L_1))$.
We note that $|(x_1(R)-x_0(R))|$ grows exponentially with $R \geq R_0$.

Having defined $S_z(R)$ as an ordered sequence of $ r \in \mathbb{Z}[\rho]$ such that $ r \in (x_0(R'),x_1(R'))$ for some $R' \in [R_0,R]$, we conclude that the cardinality of $S_z(R)$ grows exponentially with $R \geq R_0$.
Indeed, when $|(x_1(R)-x_0(R))|$ is exponentially large, we can subdivide it into exponentially large number of subsegments, each minimally satisfying the condition of the appropriate bullet lemma, and algorithmically find an element of the $\mathbb{Z}[\rho]$ in each subsegment.
Therefore, if an initial subsequence of any $S_z(R)$ has cardinality in $O(L_1)$, then that subsequence is also contained in some $S_z(R')$, where $R'$ is in $\log_2(|z^{\bullet}/z|)+O(\log(L_1))$.

Per Conjecture \ref{cnj:enough:normeq} it would be sufficient to inspect $O(L_1)$ initial elements of a large enough $S_z(R)$ in order to find $r \in S_z(R)$ such that the norm equation $|y|^2 = 2^{L_r} - | r z|^2$ is solvable.
By the above observation, such an $r$ will be also in some $S_z(R')$, where $R' = \log_2(|z^{\bullet}/z|)+O(\log(L_1))$, and the lemma follows.


\end{proof}

\begin{corol} \label{Lr:asymptotics}
In the context of Lemma \ref{r:sampling:lemma} one can efficiently find an $r \in \mathbb{Z}[\rho]$ such that
\begin{equation*}
L_r \leq \log_2(1/\varepsilon)/2 + O(\log(\log(1/\varepsilon))) + c,
\end{equation*}
where $c$ is a constant.
\end{corol}

\begin{proof}
We find an $r \in \mathbb{Z}[\rho]$ that satisfies the conditions of Lemma \ref{r:sampling:lemma}.
Recall that up to an additive fractional part $L_1=2   \log_2(|z|)$.
By condition 1, $L_r \leq \log_2(|z|) + \log_2(|z^{\bullet}|)+O(\log(\log(|z|))$.
As per Corollary \ref{corol:integer:rel:general} and Observation \ref{the:bullet:observ}, both $|z|$ and $|z^{\bullet}|$ are in $O(\varepsilon^{-1/4})$ and our claim follows.
\end{proof}

We now have an algorithm for Stage 2 that iterates through a sufficient set of candidate values of $r$ until one yields a solvable norm equation.
The pseudocode is shown in Fig.~\ref{fig:rand:normalization:one}.

\begin{figure}[t]
\begin{algorithmic}[1]
\Require $z \in \mathbb{Z}[\omega]$
\Comment hyperparameters $ \rho, \upsilon$
\Procedure{NORMALIZATION-1}{$z$}
\State $L_1 \gets \lceil \log_2(|z|^2) \rceil , \zeta \gets L_1 - \log_2(|z|^2), Y\gets None $
\State $R_0 \gets \lceil \log_2(|z^{\bullet}|/|z|    L_1   \upsilon^2) \rceil $
\State $x_{min} \gets (1-1/(2  L_1)   2^{(R_0+\zeta)/2}, x_{max} \gets  2^{(R_0+\zeta)/2}$
\State $\Delta \gets x_{max}-x_{min}$

\While{ $Y = None$ }
\State $x_0 \gets x_{min}, x_1 \gets x_0+\Delta $
\While{ $Y = None$ and $x_1 \leq x_{max}$ }

  \State construct $r \in \mathbb{Z}[\rho] , r \in (x_0,x_1)$ such that
  \State $|r^{\bullet}| \leq 2^{(R_0+\zeta)/2}$
  \State $ L_r \gets \lceil \log_2(|r  z|^2) \rceil$
  \If {$|y|^2 = 2^{L_r} - |r  z|^2 $ is easily solvable}
    \State $Y \gets \{r, y\}$

  \EndIf
  \State $x_0 \gets x_1, x_1 \gets x_1 + \Delta$
\EndWhile
\State $x_{min} \gets 2   x_{min}, x_{max} \gets 2   x_{max}$
\EndWhile
\EndProcedure
\Ensure $Y$ \Comment acceptable norm equation solution
\end{algorithmic}
\caption{\label{fig:rand:normalization:one}Algorithm to find a probability modifier $r$. }
\end{figure}

\subsection{Stage 3: PQF Unitary Design} \label{sec:circuit:design}
When the algorithm to modify the probability succeeds for a given $z$, we can construct a single-qubit unitary $V$ of the form~(\ref{general:z:y:matrix}), where $y, z \in \mathbb{Z}[\omega]$,   $L \in \mathbb{Z}$, $\nu=\sqrt{2}$, and the probability of success of the current round is $|z|^2/2^L > 1-1/L$.
For Clifford+$T$, the unitary $V$ can be decomposed exactly into an optimal ancilla-free Clifford+$T$ circuit using methods in \cite{KMM12}.
For Clifford+$\pi/12$, we decompose it using a similar technique described in Appendix \ref{app:exact:pi12}.

The following theorem summarizes the theoretical upper bounds on the mean expected cost of a PQF circuit over the Clifford+$T$ and Clifford+$\pi/12$ bases.
For completeness, we highlight that the same bound in fact applies to RUS circuits over the Clifford+$T$, resulting in a small but definitive asymptotic improvement over the bound given in Ref.~\cite{BSRRUS}.

\begin{thm} \label{corol:expected:cost:bound}
In the context of both PQF and RUS protocols where modifier sampling is based on Lemma \ref{r:sampling:lemma}
\begin{enumerate}
\item For the Clifford+$T$ basis, if the $T$-cost of the fallback round of PQF is in $O(\log(1/\varepsilon))$ then the overall expected $T$-cost of a one-round PQF protocol is
\begin{equation} \label{expected:cost:bound}
\log_2(1/\varepsilon)+O(\log(\log(1/\varepsilon))),
\end{equation}

\item The expected $T$-cost of an RUS protocol is also given by Eq (\ref{expected:cost:bound}).

\item For the Clifford+$\pi/12$ basis, the overall expected $K$-cost of a one-round PQF protocol is
\begin{equation} \label{expected:cost:bound:pi12}
1/2   \log_2(1/\varepsilon)+O(\log(\log(1/\varepsilon))),
\end{equation}

\end{enumerate}
\end{thm}

\begin{proof}

\begin{enumerate}
\item For the PQF protocol over the Clifford+$T$ basis, the expected $T$-cost is $2  L_r + C_F(1-p(r))$, where $C_F$ is the fallback cost.
As per the above and Observation \ref{observe:size:lower:T},  $1-p(r) < 2/\log_2(1/\varepsilon)$ and by Corollary \ref{Lr:asymptotics}, the claim follows.

\item For the RUS protocol with Lemma \ref{r:sampling:lemma}, the expected $T$-cost is $2   L_r /p(r)$.
As per condition 2 of the lemma, the expected cost is dominated by $2   L_r(1+1/L_1)   =  2   L_r(1+2/\log_2(1/\varepsilon))  $ and the claim follows from Corollary \ref{Lr:asymptotics}.

\item For the PQF protocol over the Clifford+$\pi/12$ basis, the expected $K$-cost is bounded by  $ L_r + 2 + C_F(1-p(r))$, where $C_F$ is the fallback cost.
As per the above
and Observation \ref{observe:size:lower:T},
$1-p(r) < 2/\log_2(1/\varepsilon)$ and by Corollary \ref{Lr:asymptotics} the claim follows.
\end{enumerate}

\end{proof}

\subsection{Stage 4: Synthesis of PQF Subcircuit}
From the unitary matrix $V$, we construct a two-qubit unitary $U$ given by
\begin{equation*}
U = \mbox{CNOT}   (I \otimes V)  \mbox{CNOT} =  \left[\begin{smallmatrix}
              V & 0 \\
              0 & X  V   X
            \end{smallmatrix}\right].
\end{equation*}
We denote the primary input state for round $k$ as $|\psi_k\rangle$.
The subcircuit $U$ for round $k$ acts on the state $\ket{\psi_k}\otimes\ket{0}$, where the second qubit is an ancilla.
We then measure the second (ancilla) qubit.

When the measurement outcome is $0$, the first qubit is left in the state
$\left[\begin{smallmatrix}
              1 & 0 \\
              0 & z^*/z
            \end{smallmatrix}\right]   |\psi_k\rangle$
which is the desired $\varepsilon$-approximation of $R_z(\theta)$.

When the measurement outcome is $1$, the first qubit is left in the state
$\left[\begin{smallmatrix}
              1 & 0 \\
              0 & -y/y^*
            \end{smallmatrix}\right]   |\psi_k\rangle$.
Unless $-y/y^*$ is $\varepsilon$-close to $e^{i   \theta}$, in this case we must apply the rotation $R_z(\theta')$, where $\theta' = \theta - \arg(-y/y^*)$ in the next round.


The unitary $U$ at round $k$ has the same $T$-count ($K$-count) as
 the $T$-count ($K$-count) of the optimal single-qubit Clifford+$T$ (Clifford+$\pi/12$) circuit for unitary $V$ since we invoke the optimal single-qubit deterministic decomposition of $V$ to obtain its circuit.
The only other gates involved are two $\mbox{CNOT}$ gates.
For Clifford+$T$, the techniques in Refs.~\ref{Selinger,RoSelinger,KMM1231,Kliuchnikoff} can be used to optimally decompose $V$.
In Appendices \ref{app:exact:pi12}--\ref{app:appox:rt3}, we show how to optimally decompose a single-qubit gate into the Clifford+$\pi/12$ basis. 
The $T$-count ($K$-count) of the two-qubit unitary at any subsequent round is defined (asymptotically) by the precision $\varepsilon$.
The difference in cost between the rounds is asymptotically bounded by an $O(\log(\log(1/\varepsilon)))$ term.

\section{Numerical Results} \label{sec:experimental:results}

We evaluate the performance of our algorithm on a set of $1000$ angles randomly drawn from the interval $(0,\pi/2)$ at $30$ target precisions $\varepsilon \in \{10^{-11}, \ldots, 10^{-40}\}$. 
In all numerical experiments, expected cost statistics have been collected for one-round PQF circuits.
Adding the second round to the compiled circuits only improves the mean expected gate count by $3$ gates on average. 
This is due to the probability modification at Stage 2 of per-round PQF compilation.
Modification boosts the probability of success to typical values above $0.97$ and above $0.985$ for at least half of the cases.
%
%

\begin{figure}[bt]
\includegraphics[width=3.5in]{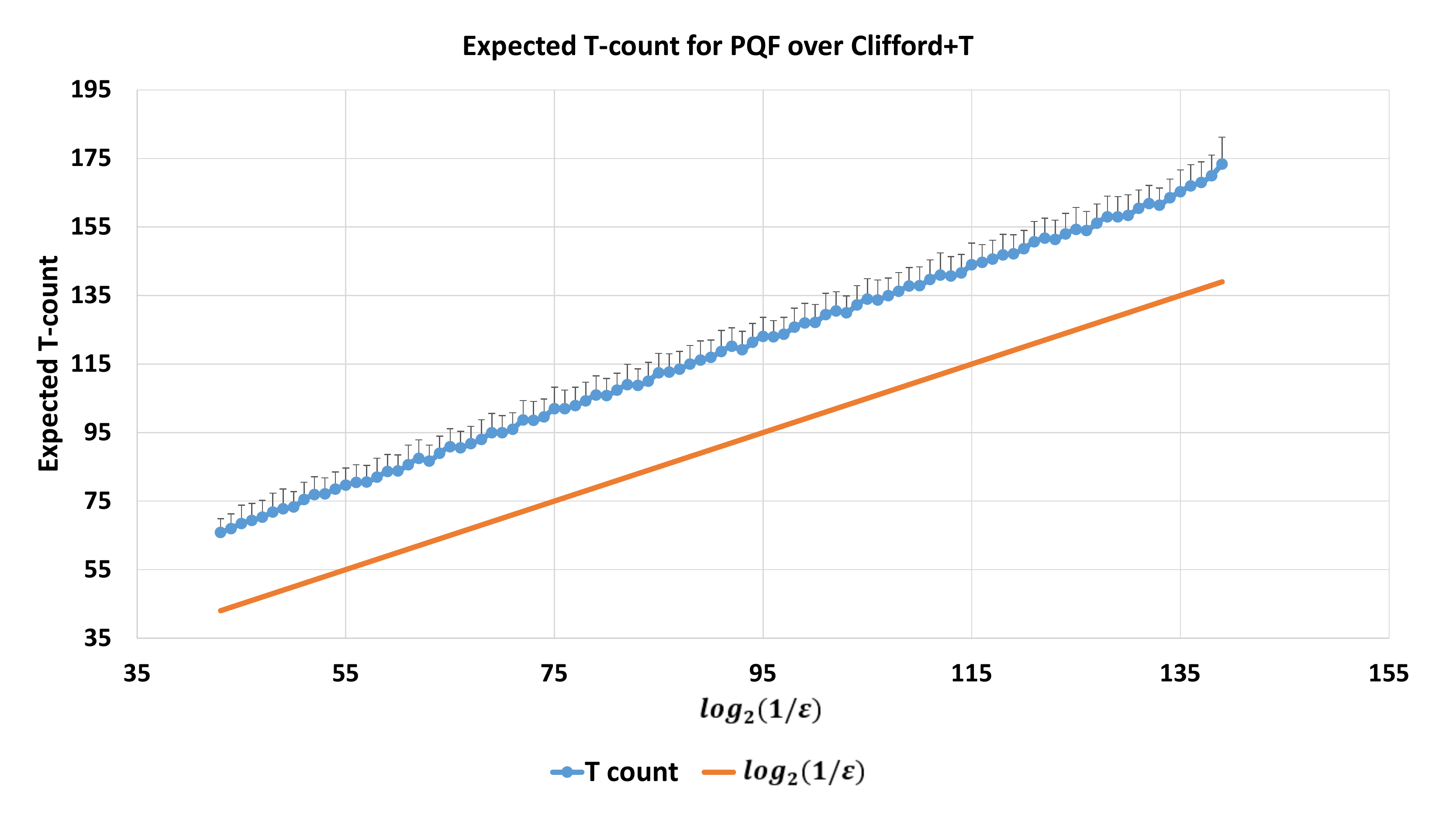}
\caption{\label{fig:random:angles} Precision $\varepsilon$ versus mean expected $T$-count of PQF circuits for the set of random angles. } 
\end{figure}

Figure \ref{fig:random:angles} plots the precision $\varepsilon$ versus the mean (and standard deviation) of the expected $T$-count across the PQF circuits generated for the set of $1000$ random angles.

The maximum likelihood estimate for the mean expected $T$-count is $\log_2(1/\varepsilon)+4\,\log_2(\log_2(1/\varepsilon))+1.187$.

\begin{figure}[bt]
\includegraphics[width=3.5in]{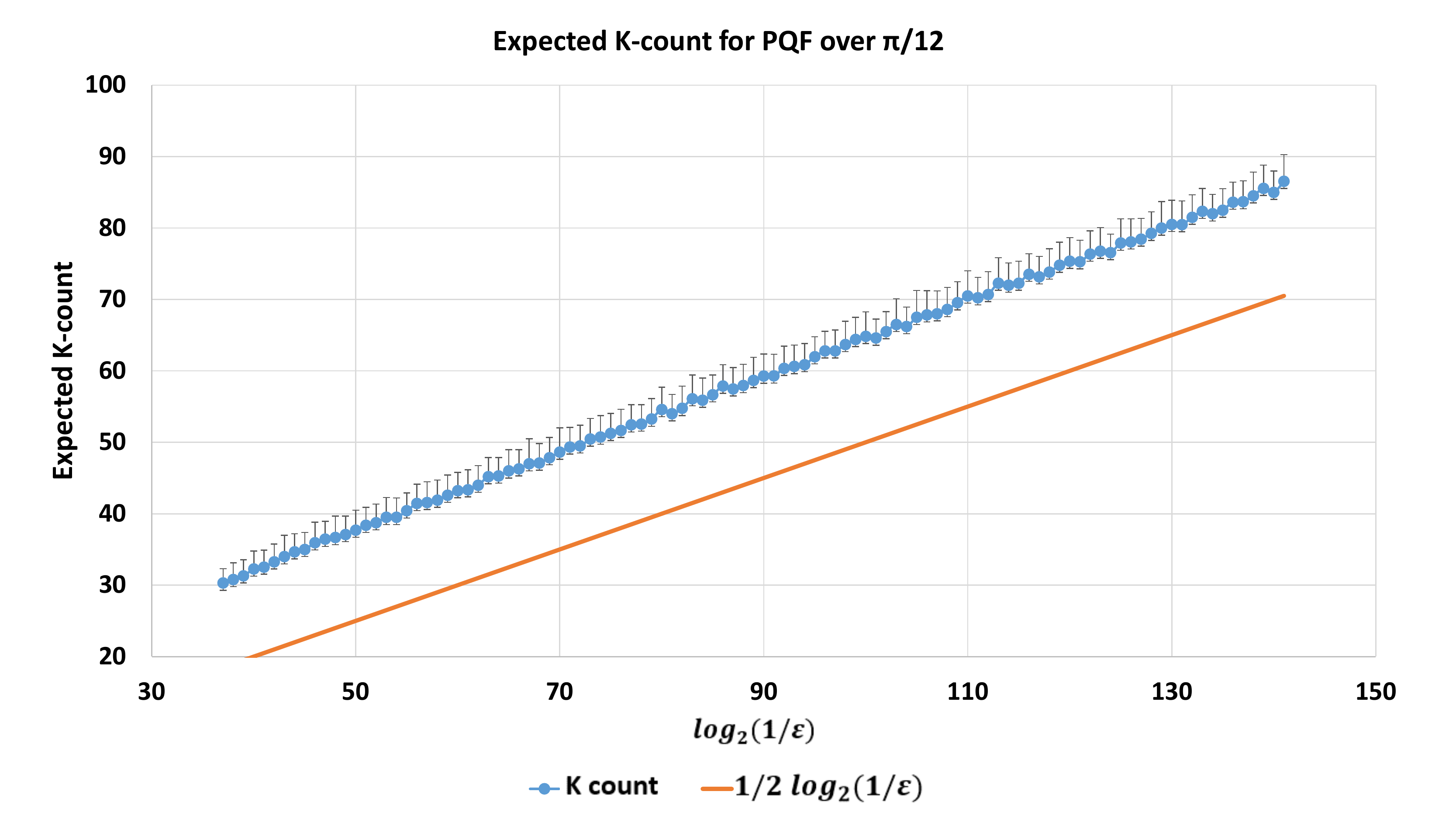}
\caption{\label{fig:pi12:random:angles} Precision $\varepsilon$ versus mean expected $K$-count of PQF circuits over Clifford+$\pi/12$  for the set of random angles. } 
\end{figure}

Figure \ref{fig:pi12:random:angles} plots the precision $\varepsilon$ versus the mean (and standard deviation) of the expected $K$-count for the PQF Clifford+$\pi/12$ circuits generated for the set of $1000$ random angles.
The maximum likelihood estimate for the mean expected $K$-count is $1/2 \log_2(1/\varepsilon)+2\log_2(\log_2(1/\varepsilon))+3.48$.

\begin{figure}[bt]
\includegraphics[width=3.5in]{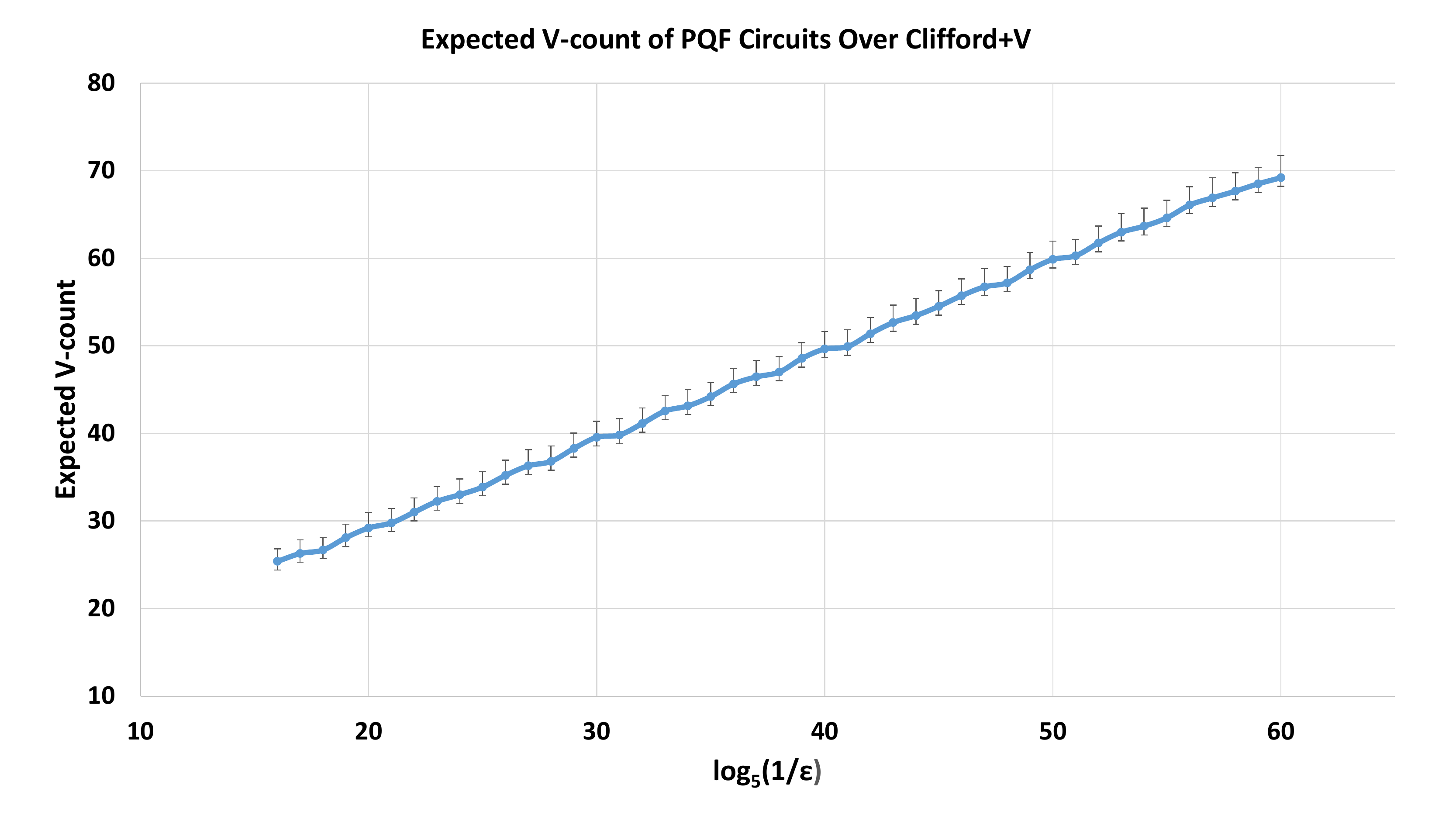}
\caption{\label{fig:V:random:angles} Precision $\varepsilon$ versus mean expected $V$-count of PQF circuits over Clifford+$V$  for the set of random angles. } 
\end{figure}

Figure \ref{fig:V:random:angles} plots the precision $\varepsilon$ versus the mean (and standard deviation) of the expected $V$-count for the PQF Clifford+$V$ circuits generated for the set of $1000$ random angles.
The maximum likelihood estimate for the mean expected $V$-count is $\log_5(1/\varepsilon)+0.95\log_5(\log_5(1/\varepsilon))+7.26$.

\section{Conclusion and Future Work}
\label{sec:conclude}

We have developed a method of synthesizing Probabilistic Quantum Circuits with Fallback (PQF) which is simpler and more general than the synthesis of Repeat-Until-Success (RUS) circuits. 
We have demonstrated that the method can be applied to the approximation of single-qubit unitaries over at least three different universal quantum bases. 
The mean expected cost of the resulting probabilistic circuits has an asymptotic upper bound with a leading term that is $3$ times smaller than the leading term of the corresponding optimal, purely unitary, ancilla-free circuit over the same universal quantum basis.
In addition, our PQF protocol requires only a small finite number of steps to achieve efficient approximation of a unitary.

The design and cost analysis of PQF circuits is performed based on conjectures that are remarkably similar to the norm density conjectures presented in Ref.~\cite{Selinger} and Ref.~\cite{RoSelinger}. 
Our numerical experiments, covering around $30,000$ synthesis instances for each of the three universal quantum bases, have not produced a single instance that would violate our underlying conjectures.

In contrast to the RUS protocol, the PQF protocol is remarkably general, and promises generalization to multi-qubit unitary decomposition and synthesisi. 
Future work includes generalizing PQF to multi-qubit target unitaries, and to additional universal bases, most notably to Clifford+$\pi/16$.
For each basis considered, it is important to also determine a fault-tolerant implementation of the non-Clifford gates employed.
For example, for Clifford+$\pi/12$, it will be important to construct either distillation methods or circuit constructions for a fault-tolerant $\pi/12$ gate.
Recent methods have addressed the distillation of non-Clifford states and Fourier states, and provide a starting point for research \cite{DCP,DCS} on other possible universal bases and their fault-tolerant constructions.
Consideration of generalization to qudit computation models is also an avenue for future exploration.
Finally, formal, rigorous proofs of the underlying conjectures is another important direction.



%

\appendix

\section{Information-Theoretic Bounds} \label{sec:theoretic:bounds}

A relatively simple analysis of the density with which cyclotomic rationals are distributed imposes information-theoretic limits on how much we can reduce the expected $T$-count of our non-deterministic solutions compared to the $T$-count of deterministic, unitary solutions.
Note that the analysis applies equally well to both PQF and RUS methods.

Let us assume, temporarily, that for $z \in \mathbb{Z}[\omega]$ and $L = \lceil \log_2(|z|^2) \rceil$, the norm equation $|y|^2 = 2^L - |z]^2$ is solvable.
By definition of $L$, $|z|^2 \leq 2^L$.
We know that the optimal $T$-count of a single-qubit unitary circuit implementing a matrix of the form of Eq~(\ref{general:z:y:matrix}) is $t = 2 L$ or $t = 2 L -2$.

In either case we note that $|z|^2 = O(2^{t/2})$ and $|z|^4 = O(2^t)$.
We also note that given an upper bound $b$ on the absolute value of cyclotomic integer, there are no more than $O(b^4)$ cyclotomic integers under this bound.
Thus we conclude that there are no more than $O(2^t)$ cyclotomic integers $z$ for which the matrix of the form of Eq~(\ref{general:z:y:matrix}) may exist and be implemented at $T$-count $t$ or less.

It follows that there are at most $O(2^t)$ unimodular cyclotomic rationals on the unit circumference for which our RUS circuit can be built with design cost of $T$-count $=t$ or less.
Therefore, there exists a constant $K$ such that for $\varepsilon < K \times 2^{-t}$ there is an arc of the unit circumference of length $2   \varepsilon$ that does not contain any such cyclotomic rational.
If $\theta_*$ is the angle in the center of such an arc, then the rotation $R_z(\theta_*)$ cannot be implemented by any of our circuits with design cost of $T$-count $=t$ or less.

Conversely, $\varepsilon \geq  K \times 2^{-t}$ is the necessary condition for any axial rotation to be implementable by one of our circuits with design cost of $T$-count $=t$ or less.
This necessary condition is equivalent to
\begin{equation} \label{eq:RUS:lower:bound}
t \geq \log_2(1/\varepsilon) + \log_2(K),
\end{equation}
which is a specific lower bound on the design cost given by the $T$-count of our solution.

The derivation of the above lower bound is specific to our PQF and RUS designs.
As follows from Corollary \ref{corol:expected:cost:bound}, our existing PQF protocol for single-qubit decomposition based on the PSLQ integer relation algorithm is within $O(\log(\log(1/\varepsilon)))$ from this bound.
Deriving a uniform lower bound under more general assumptions would be a worthwhile problem for future research.

\section{Details on the Norm Equation in $\mathbb{Z}[\omega]$}
\label{app:norm}

This section combines the claims for $\omega=e^{2   \pi   i  /m}$ for $m=8$ and $m=12$.

We reintroduce $\rho=\omega+\omega^*$.
Recall that the real-valued ring $\mathbb{Z}[\rho]$ is a unique factorization ring.
That is, any of its elements can be factored into a product of prime algebraic integers and at most one unit.
The primary category of right-hand-side values for which Eq~(\ref{eq:norm:equation}) is easily solvable would then be the set of algebraic integer primes.

The equation is easily solvable for the following kinds of prime right-hand sides  (c.f.~\cite{LWashington}):
\begin{enumerate}
\item $\xi = a + b   \rho$, $\xi > 0$  and $p=\xi   \xi^{\bullet}$ is a positive rational prime number with $p = 1   \mod   m$;
\item $\xi$ is a rational prime number and $\xi \neq -1   \mod   m$.
\end{enumerate}
We call an algebraic integer prime belonging to one of these two classes a ``good'' prime.

For a composite $\xi$ we consider a \emph{limited} factorization of the right-hand side to preserve efficiency.
To this end, we precompute a set $S_{prime} \subset \mathbb{Z}[\rho]$ of small prime elements and consider factorizations of the form:
$\xi =  \xi_1^{a_1}   \ldots, \xi_r^{a_r}   \eta$, where $\xi_1, \ldots, \xi_r \in S_{prime}$ and $\eta$ passes a primality test.
Eq~(\ref{eq:norm:equation}) is efficiently solvable if $\eta$ is a good prime and for $i=1, \ldots , r$, $\xi_i$ is a good prime or $a_i$ is even.

\begin{example}
For $m=8$,
$|y|^2 = \xi = 1270080 + 211680   \sqrt{2}$ is efficiently solvable since
$\xi= 2^5   3^3   5   7^2   (2+\sqrt{2}) (5- 2  \sqrt{2})$.
\end{example}

Note $p= (5- 2  \sqrt{2})  (5- 2  \sqrt{2})^{\bullet}= 17 = 1   \mod   8$.
The only ``bad" prime in the above factorization is $7$ but it appears as an even power.

We remark that the cyclotomic integer $z$ coming from the cyclotomic rational approximation of
$e^{i   \theta}$ is not unique.
In fact, it is defined up to an arbitrary real-valued factor $r \in \mathbb{Z}[\rho]$.
For any such $r$, $(r  z)^* / (r   z)$  is identical to $z^*/z$.
However the norm equation $|y|^2 = 2^L - |r   z|^2$ can and will change quite dramatically.

When drawing $r$ randomly from a subset of $\mathbb{Z}[\rho]$ one might try and estimate the chance that the equation $|y|^2 = 2^L - |r   z|^2$ turns out to be solvable for a random $r$.
This is an example of an open and likely very hard number theory problem.
We will not attempt to solve it here and will instead rely on a conjecture that the ``lucky" values of $r$ are reasonably dense in $\mathbb{Z}[\rho]$.

\section{Exactly Representable Single-Qubit Circuits in Clifford+$\pi/12$} \label{app:exact:pi12}

We use the notation $\omega=\omega_{12}=e^{i \,  \pi/6}$ in this section.

In this and subsequent sections we also use a shorthand notation for single-qubit controlled phase gate. Given $\phi \in \mathbb{C}, |\phi|=1$ is a phase factor, the \emph{controlled phase gate} $\Lambda(\phi)$ is simply $\left[\begin{smallmatrix}1&0\\0&\phi\end{smallmatrix}\right]$.

In particular, the $\pi/12$ gate $K=\Lambda(\omega_{12})$.

The single-qubit Clifford+$\pi/12$ group is generated by the Hadamard gate $H$ and the $\pi/12$ gate $K$.
We note that $\omega^3=i$ and therefore the common phase gate $S=K^3$ is in the circuit group, as is, by closure, the entire single-qubit Clifford group.

Any Clifford+$\pi/12$ circuit can be expressed as a product of syllables of the form $K^k   H$, where $|k| < 6$, up to a possible global phase factor.
A slightly deeper analysis reveals that we can rewrite a circuit to enforce $k={\pm 1, \pm 2}$ in all  interior syllables, but this is not very important in this section.
The important part is that the $K^k   H$ syllable is a Clifford gate for $k \in \{0, \pm 3, \pm 6, \pm 9\}$ and has zero $K$-count.

We assume that the implementation cost of gates of the form  $K^k$, $k \notin \{0, \pm 3, \pm 6, \pm 9\}$  is the same and that it is significantly higher than the cost of a Clifford gate.
This implies that the $K$-cost of a circuit composed of $K^k   H$ syllables is upper-bounded by the number of syllables with $k \notin \{0, \pm 3, \pm 6, \pm 9\}$.

Consider the ring of cyclotomic integers $\mathbb{Z}[\omega]$.
Any $\omega^k$ is a cyclotomic integer and $H=\frac{1}{\sqrt{2}}   \left[\begin{smallmatrix}1&1\\1&\textrm{-}1\end{smallmatrix}\right]$, where $\pm 1$ are in $\mathbb{Z}[\omega]$ and  $\frac{1}{\sqrt{2}}$ is not in  $\mathbb{Z}[\omega]$.
Clearly a finite product of the $K^k   H$ syllables evaluates to a unitary matrix of the form

\begin{equation} \label{pi12:z:y:matrix}
\frac{1}{{\sqrt{2}}^L}   \left[\begin{array}{cc}
              z & -y^*   \omega^\ell \\
              y & z^*   \omega^\ell
            \end{array}\right],
\end{equation}
$y, z \in \mathbb{Z}[\omega]$, $\ell, L \in \mathbb{Z}$.

\begin{lem} \label{lem:zero:L}

A Clifford+$\pi/12$ circuit that evaluates to a unitary in the form (\ref{pi12:z:y:matrix}) with $L = 0$ has $K$-cost $0$ or $1$.
\end{lem}

\begin{proof}
The unitarity of (\ref{pi12:z:y:matrix}) with $L = 0$ means $|z|^2+|y|^2=1$. Since $y,z$ are algebraic integers, either $|z|=1, |y|=0$ or  $|z|=0, |y|=1$.
By standard algebraic units argument, if $x \in \mathbb{Z}[\omega]$ and $|x|=1$ then,  $x=\omega^k, k \in \mathbb{Z}$.

In the case $|z|=1$ and $z=\omega^k$, the unitary in the form (\ref{pi12:z:y:matrix}) is $\omega^k  \Lambda(\omega^{\ell-2   k})$. As per the assumptions we have adopted above, the $\pi/12$-cost of the latter is either $0$ or $1$.

The case of $|z|=0$ is reduced to the case of $|z|=1$ by pre-multiplying the subject unitary times $X=H  Z   H = H  \Lambda(\omega^6)   H$. By our convention the latter has zero $K$-cost and does not affect the $K$-cost of the resulting circuit.
\end{proof}

\begin{lem} \label{lem:column:in:L}
Consider a unitary $2$-vector of the form $v=\frac{1}{{\sqrt{2}}^L}   (z,y)^T$,  $y, z \in \mathbb{Z}[\omega]$, $L \in \mathbb{Z}$ : $|y|^2+|z|^2=2^L$.
A Clifford+$\pi/12$ circuit $c$ with $K$-cost at most $L+1$ can be algorithmically found such that $c \,  v = (1,0)^T$.

\end{lem}

\begin{proof}
This rather technical lemma is inspired by the ``column lemma" from \cite{GilSel}.
The proof is by induction in $L$.
The base of the induction is $L=0$, and the claim has been already established in the proof of Lemma \ref{lem:zero:L}.

Consider the subject vector with $L>0$.
The main step is to algorithmically find a short circuit $c$ with $K$-cost at most $1$ such that $v'=c  \,  v$ is or the form $v'=\frac{1}{{\sqrt{2}}^{L'}}   (z',y')^T $, where $L' < L$.
Then the desired circuit will be generally of the form $H   \Lambda(\omega^k)$, except for one special case where it will be a global phase.
We generally attempt to find  $k \in \mathbb{Z}$ such that all the integer coefficients of the algebraic integers $z\pm\omega^k  y$ are even.
If we have succeeded in finding such a $k$ then
\begin{eqnarray*}
H   \Lambda(\omega^k)   v &=& \frac{1}{{\sqrt{2}}^{L+1}}   (z+\omega^k  y,z-\omega^k  y)^T \\
&=& \frac{1}{{\sqrt{2}}^{L-1}}   ((z+\omega^k  y)/2,(z+\omega^k  y)/2)^T
\end{eqnarray*}
and we have succeeded in reducing the denominator exponent.

In order to develop a method for finding the desired $k$, consider the parity morphism
\begin{eqnarray*}
\mu&:& \mathbb{Z}[\omega] \rightarrow \mathbb{Z}_2[\omega]  \\
\mu&:& a   \omega^3 + b   \omega^2 + c   \omega + d  \\
&\mapsto& (a\, \mbox{mod} 2)   \omega^3 + (b\, \mbox{mod} 2)  \omega^2 + (c\, \mbox{mod} 2)   \omega + (d\, \mbox{mod} 2).
\end{eqnarray*}

All coefficients of $z\pm\omega^k  y$ are even if and only if  $0=\mu(z\pm\omega^k  y)=\mu(z) \oplus \omega^k  \mu(y)$  if and only if $\mu(z)= \omega^k  \mu(y)$, which is going to be the desired property below.

Consider the action of the 12-element group $\{\omega^k\}$ on $\mathbb{Z}_2[\omega]$ by multiplication.
Since $\omega^6=-1$ and $-1 = 1 \mod 2$ the subgroup $\{1,-1\}$ acts trivially on $\mathbb{Z}_2[\omega]$ and the action of the 6-element factor-group $\{\omega^k\}/\{1,-1\}$ is well-defined.

By direct computation we established that the the 16-element set $\mathbb{Z}_2[\omega]$ is partitioned into $4$ orbits of this action.
The orbit of zero $O_0$ consists of just zero.
The orbit of $1+\omega^3=1+i$ ,  $O_3$ consists of $3$ elements and the orbits $O_1$ and $O_2$ of $1$ and $1+\omega$ respectively consist of $6$ elements each.

It is important to understand that the function $N_2 : x \mapsto \mu(|x|^2)$ is constant on each of the orbits.
More specifically, $N_2(O_0)=N_2(O_3)=0$ , $N(O_1)=1$, $N(O_2)=\omega^3$.
The final key remark is that the unitarity of the vector $v$ implies $N_2(z)\oplus N_2(y)=N_2(2^L)=0$ and therefore $N_2(z)=N_2(y)$.

We proceed by case distinction.

(0,0) Case of $O_0$

If both $\mu(z)$ and  $\mu(y)$ belong to $O_0$ then all the integer coefficients of $y$ and $z$ are already even and we do not need to do any transformations in order to reduce the vector.

(1,2) Cases of $O_1$ and $O_2$

If $\mu(z)$ belongs to either of the two orbits, then $\mu(y)$ must belong to the same orbit (since we have established $N_2(z)=N_2(y)$). Therefore there exists $k$ such that $\mu(z)= \omega^k  \mu(y)$ which is what we were looking for.

(3,3) Case of $O_3$

If both $\mu(z)$ and  $\mu(y)$ belong to $O_3$, then again there exists $k$ such that $\mu(z)= \omega^k  \mu(y)$

(3,0) This is the only remaining case.

If one and only one of the $\mu(z)$,  $\mu(y)$ belongs $O_3$ then the other one must belong to $O_0$ (since these are the only two orbits with $N_2(orbit)=0$).
Assume, w.l.o.g. that $\mu(y)=0$.

This case needs to be treated differently from the general context. First, we note that, since $\mu(z) \in O_3$ there exists a $k$ such that $\mu(\omega^k   z)=1+\omega^3=1+i$. Next we note that the global phase operator $(1+i)/\sqrt{2}   I_2$ is in the the Clifford group and that $\mu((1+i)^2)=\mu(2  \omega^3)=0$.
Therefore by multiplying the vector $v$ times the global phase $\omega^k   (1+i)/\sqrt{2}   I_2$ we obtain a vector, where all the integer coefficients of both components are even.
We then reduce this latter vector to one of the form $1/{\sqrt{2}}^{L-1}   w$.

This case concludes the induction step.

\end{proof}

\begin{corol} \label{pi12:corol:decomposition}

Unitary of the form (\ref{pi12:z:y:matrix}) where $y, z \in \mathbb{Z}[\omega], L , k\in \mathbb{Z}$ can be represented exactly and algorithmically by a Clifford+$\pi/12$ circuit of $\pi/12$-count at most $L+2$.

\end{corol}
\begin{proof}
Consider a Clifford+$\pi/12$ circuit $c$ of $\pi/12$-count at most $L+1$ that reduces the fist column of the matrix (\ref{pi12:z:y:matrix}) to $(1,0)^T$. Consider the unitary value of $c^{\dagger}= c^{\dagger}   I_2$. Since $c^{\dagger}$ maps $(1,0)^T$ into the first column of (\ref{pi12:z:y:matrix}) it maps $(0,1)^T$ into a unitary vector that is Hilbert-orthogonal to that first column. Thus $c^{\dagger}   (0,1)^T$ is proportional to $\frac{1}{{\sqrt{2}}^L}  (-y^*   \omega^\ell ,  z^*   \omega^\ell)^T$ with a unit coefficient from
$\mathbb{Z}[\omega]$. Therefore we can algorithmically find an integer $k$ such that the unitary (\ref{pi12:z:y:matrix}) is exactly equal to the value of  $c^{\dagger}   \Lambda(\omega^k)$. Since the $\pi/12$-count of the $\Lambda(\omega^k)$ is at most $1$ the corollary follows.
\end{proof}

\section{Approximation of Real Numbers by Numbers from $\mathbb{Z}[\sqrt{3}]$} \label{app:appox:rt3}

This section is a direct extension of Section 5 in \cite{Selinger} to the $\mathbb{Z}[\sqrt{3}]$ ring.
Recall that the fundamental Galois automorphism of that ring extends  $\bullet: \sqrt{3} \mapsto (- \sqrt{3})$.
The following is an analog of Lemma 17 from \cite{Selinger}:

\begin{lem} \label{sqrt:3:approximation}
Let $[x_0,x_1]$ and $[y_0,y_1]$ be closed intervals of real numbers. Let $\delta=x_1-x_0$ and $\Delta=y_0-y_1$, and assume $\delta   \Delta \geq (2+\sqrt{3})^2$. Then there exists at least one $\alpha=a+b \sqrt{3} \in \mathbb{Z}[\sqrt{3}]$ such that $\alpha \in [x_0,x_1]$ and $\alpha^{\bullet} = a- b \sqrt{3} \in [y_0,y_1]$. Moreover, there is an efficient algorithm for computing such $a$ and $b$.
\end{lem}

The proof is almost identical to the proof of the lemma for $\mathbb{Z}[\sqrt{2}]$ with the obvious replacements of $\sqrt{2}$ by $\sqrt{3}$ and of the unit $\lambda=1+\sqrt{2}$ by the unit $\upsilon=2+\sqrt{3}$.

\section{Approximating Single-Qubit Circuits in Clifford+$\pi/12$}  \label{app:appox:pi12}

This section is a direct extension of Section 6 in \cite{Selinger} to the $\mathbb{Z}[\omega=e^{i   pi/6}]$ ring.
We prove that an axial rotation $\Lambda(e^{i   \theta})$ can be algorithmically approximated to any desired precision $\varepsilon>0$ by a Clifford+$\pi/12$ circuit with $\pi/12$-count of at most $2   \log_{\sqrt{2}}(1/\varepsilon)+C$, where $C=3/2 + \log_{\sqrt{2}}(2+\sqrt{3})$.

Recall that $\sqrt{3}=2  \omega - \omega^3$ and $i=\omega^3$ and consider the subring $\mathbb{Z}[\sqrt{3}][i] \subset  \mathbb{Z}[\omega]$.
Let $\theta \in \mathbb{R}$ and $\varepsilon >0$  be fixed.

\begin{defin}  \label{def:feasible:candidate}

Consider some  $u = (a+b  \sqrt{3}) + ((c+d  \sqrt{3})  i \in \mathbb{Z}[\sqrt{3}][i]$.
Complex number $u/{\sqrt{2}}^k,   k \in \mathbb{Z}$ is called a \emph{feasible candidate at round $k$} for $(\theta,\varepsilon)$ if
\begin{enumerate}
\item $|u^{\bullet}|^2 \leq 2^k$;
\item $|u|^2 \leq 2^k$ and $Re(u   e^{i   \theta/2}) \geq (1-\varepsilon^2)  {\sqrt{2}}^k$.
\end{enumerate}
\end{defin}

\begin{thm} \label{thm:feasible:candidates}

Let $\varepsilon >0$ and $\theta \in \mathbb{R}$ be fixed and let $k \geq C+  log_{\sqrt{2}}(1/\varepsilon)$, where $C=1/2 + \log_{\sqrt{2}}(2+\sqrt{3})$.
Then there exists a set of at least $n=\lfloor 2  \sqrt{2}/\varepsilon \rfloor$ feasible candidates at round $k$ for $(\theta,\varepsilon)$.
Moreover there is an efficient algorithm for generating a sequence of random candidates from this set.
\end{thm}

\begin{proof}
First note that $k \geq C+  log_{\sqrt{2}}(1/\varepsilon)$  implies $2^k \geq  \sqrt{2}    (2+\sqrt{3})^2/\varepsilon^2$.
Define
$\delta = {\sqrt{2}}^k   \varepsilon^2$ and $\Delta = {\sqrt{2}}^{k+1}$.
and observe that $\delta \Delta \geq (2+\sqrt{3})^2$ so that the criterion of Lemma \ref{sqrt:3:approximation} is satisfied for $(\delta ,\Delta)$.

For convenience we assume w.l.o.g. that
$-\pi/2 \leq \theta \leq \pi/2$.

Using the same geometric argument as in proof of Theorem 22 in \cite{Selinger} we observe that condition (2) of Definition \ref{def:feasible:candidate} defines a meniscus shape $R_{\varepsilon}$  on the complex plane. If we parameterize the plane with $x+y   i, x,y \in \mathbb{R}$ we observe that there is a vertical segment $[y_{min},y_{max}]$ such that
$y_{max} - y_{min} \geq \sqrt{2}   \varepsilon$ and such that for any $y' \in [y_{min},y_{max}]$ the intersection of the horizontal line $\{x+y'   i\}$ with the meniscus $R_{\varepsilon}$ is a segment of length at least $\varepsilon^2/2$.

Introduce $n=\lfloor 2  \sqrt{2}/\varepsilon \rfloor$
We now partition the segment $[y_{min},y_{max}]$ at points $y_j = j/n\,(y_{max} - y_{min})+y_{min}$, $j=0,\ldots,n$.
By design $y_{j+1}-y_j > \varepsilon^2/2$.

Consider closed subintervals $I_j = [y_j, y_j+\varepsilon^2/2]$, $j=0, \ldots , n-1$ that are non-overlapping subintervals of the  $[y_{min},y_{max}]$.
First we find $\beta_j \in \mathbb{Z}[\sqrt{3}]$ such that  $\beta_j \in [\sqrt{2}^k   y_j, \sqrt{2}^k   (y_j+\varepsilon^2/2)] $ and  $\beta_j^{\bullet} \in [-{\sqrt{2}}^{k-1}, {\sqrt{2}}^{k-1}]$.
This can be done algorithmically because $|[-{\sqrt{2}}^{k-1}, {\sqrt{2}}^{k-1}]|  \sqrt{2}^k   \varepsilon^2/2 \geq (2+\sqrt{3})^2$.

Let $H_j = R_{\varepsilon} \cap \{y = \beta_j/\sqrt{2}^k\}$. As we have noted the length of $H_j$ is at least $\varepsilon^2/2$.
Now find $\alpha_j \in \mathbb{Z}[\sqrt{3}]$ such that  $\alpha_j \in \sqrt{2}^k   H_j$ and  $\alpha_j^{\bullet} \in [-{\sqrt{2}}^{k-1}, {\sqrt{2}}^{k-1}]$.
This can be done algorithmically  for the same reason as above.

Clearly $(\alpha_j + \beta_j   i)/\sqrt{2}^k$ is a feasible candidate at round $k$ for $(\theta,\varepsilon)$ and it is distinct from any other such candidate $(\alpha_{j'} + \beta_{j'}   i)/\sqrt{2}^k, j \neq j'$.
By randomly selecting an integer $0 \leq j < n$ without replacement, we now can algorithmically generate a unique random feasible candidate as claimed.

\end{proof}

We now discuss a conjecture regarding solvability of a norm equation that is needed for expanding a feasible candidate $z/\sqrt{2}^L, z \in \mathbb{Z}[\omega]$ into a unitary matrix of the form (\ref{pi12:z:y:matrix}).
Such expansion exists if and only if the norm equation $|y|^2 = \xi = 2^L - |z|^2$ can be solved for $y \in \mathbb{Z}[\omega]$. In Section \ref{app:norm} we have defined the notion of easily solvable norm equation and also built up sufficient intuition the the effect that easily solvable norm equations are not uncommon. They are more common than the prime numbers with the additional property $p = 1 \mod 12$ among the integers. It is well known that in a segment of the form $[B/2,B]$ , where $B$ is sufficiently large, the density of such prime numbers is in $\Omega(1/\ln(B))$.

Further steps in the single-qubit circuit synthesis rely on the following conjecture (of the type that is now becoming common in circuit synthesis):
\begin{cnj} \label{pi12:norm:eq:conjecture}
For small enough values of $\varepsilon>0$ and $L$ in $\Omega(\log(1/\varepsilon))$ it suffices to inspect $O(\ln(2^L)) = O(L)$ feasible candidates $z/{\sqrt{2}}^L$ for
$(\theta, \epsilon)$ in order to find at least one such candidate for which the norm equation $|y|^2 = 2^L - |z|^2$ is easily solvable over $\mathbb{Z}[\omega]$.
\end{cnj}

Rigorous proof of this conjecture may be a hard number-theory problem.
At this time however we have ample numeric evidence for the conjecture for a range of $\varepsilon$ down to $10^{-100}$.

Assuming this conjecture we can claim the following:
\begin{thm}
Let $\theta$ be a fixed angle.
There exists a synthesis algorithm with probabilistically polynomial classical runtime that solves the following problem:
For a small enough value of $\varepsilon>0$ find a unitary ancilla-free Clifford+$\pi/12$ circuit with $\pi/12$-count smaller than $2 \log_2(1/\varepsilon)+K$ (where
$K=\lceil 5/2 + 2 \log_2(2+\sqrt{3}) \rceil$) that represents the axial rotation $\Lambda(e^{i   \theta})$ to absolute precision $\varepsilon$.
\end{thm}

\begin{proof}
Given $(\theta, \varepsilon)$, Theorem  \ref{thm:feasible:candidates} algorithmically defines a set of feasible candidates $z/\sqrt{2}^L$ or cardinality $\Omega(1/\varepsilon))$. For all these candidates $L \leq 2   \log_2(1/\varepsilon) + K -2$ and is in $O(\log(1/\varepsilon))$.
The outer loop of the desired algorithm randomly samples feasible candidates from the above set without replacement.
By Conjecture \ref{pi12:norm:eq:conjecture}, with arbitrarily high probability the algorithm finds a feasible candidate with an easily solvable norm equation after $O(L)$ trials.

Let $z/{\sqrt{2}}^L$ be such candidate and $y \in \mathbb{Z}[\omega]$ be a solution of the norm equation $|y|^2 = 2^L - |z|^2$. Then the unitary
\begin{equation} \label{pi12:z:y:matrix:other}
\frac{1}{{\sqrt{2}}^L}   \left[\begin{array}{cc}
              z & -y^*  \\
              y & z^*
            \end{array}\right],
\end{equation}
is an $\varepsilon$-approximation of the rotation $\Lambda(e^{i   \theta})$.
Per Corollary \ref{pi12:corol:decomposition}, this unitary can be exactly represented by a Clifford+$\pi/12$ circuit with $\pi/12$-count at most $L+2$, and the theorem follows.

\end{proof}

\section{Runtime Performance Evaluation}\label{subsec:runtime}

The PQF synthesis algorithm incurs significant compiler runtime cost at Stages 1 and 2 of each compilation round. In this section we report on some empirical findings regarding upper bounds on this compiler runtime for the various universal gate sets considered in the paper. Recall that in case of synthesis over $V$ basis the phase factor approximation is done by a simple continued fraction algorithm and its cost happens to be trivial compared to Stage 2 costs. Furthermore, in case of Clifford$+T$ and Clifford+$\pi/12$ bases, we have used a \emph{Mathematica} implementation of PSLQ algorithm published in \cite{PSLQBertok}.

The main Theorem of \cite{FergBail} states that if exact integer relations between the subject real values exist  and $M$ is the minimum norm of such an integer relation then the PSLQ algorithm terminates after a number of integration bounded by $O(\log(M))$.
Both the Theorem and the proof can be modified to apply to our customization of the algorithm that looks for approximate integer relations, to state that if $M_{\varepsilon}$ is the minimum size of an integer vector $a$ such that $|a \, x| < \varepsilon$ then the modified algorithm terminates after a number of iterations bounded by $O(\log(M_{\varepsilon}))$.
Since in the case of both $T$ and $\pi/12$ bases, $M_{\varepsilon} = O(\varepsilon^{-1/4})$,  the bound on the number of iterations to termination is linear in $\log(1/\varepsilon)$.

The Bertok implementation of the PSLQ algorithm (\cite{PSLQBertok}) appears to be asymptotically optimal in this sense.
In our experiments using $1000$ random target angles, the number of PSLQ iterations scaled on average like $1.16\,\log_2(1/\varepsilon)$ for the Clifford+T basis and scaled on
average like $1.04\,\log_2(1/\varepsilon)$ for the Clifford+$\pi/12$ basis. The standard deviation on the number of iterations computed across the test set is very moderate for both bases scaling roughly like $\log_2(\log_2(1/\varepsilon))$.

The practical cost of the Stage 1 of the compilation in fact becomes quadratic in $\log(1/\varepsilon)$ for $\varepsilon < 10^{-15}$
 when measured in common arithmetic operations that are native on a classical computer, because the PSLQ algorithm requires variable precision floating point arithmetic with precision tightening as $O(\varepsilon)$. Once the required precision exceeds the available machine precision, software simulation of variable mantissa arithmetic becomes necessary causing one-time drop in speed and subsequent quadratic trend in compilation cost.
 
The runtime of each Stage 2 compilation pass is roughly proportional to the number of candidate modification factors evaluated until an easy solution to a suitable norm equation 
is found. Each candidate factor is generated by an appropriate enumerator, then the corresponding norm equation is tested for easy solvability. The cost of generating a candidate is
trivial compared to the cost of analyzing the norm equation. The latter cost is in principle similar to the cost of testing an integer for smoothness; however in our prototype 
implementation we simply relied on the Mathematica {\sc FactorInteger} function, time-constrained to one quarter of a second. The cost of analyzing the factors has been trivial compared to the cost of the factorization.

The runtime at Stage 2 of our prototype compilation round can be upper-bounded by $k/4$ seconds, where $k$ is the number of the candidate factors needed for termination. While $k$ should be expected to scale like $O(\log(1/\varepsilon))$ with precision and does so scale in a sense, we have been finding that the $k$ and hence the runtime required at Stage 2 is a strongly stochastic variable. While said runtime has been practically acceptable in all cases, outlying cases might require an order of magnitude more candidates to terminate than the typical cases.

The details are as follows: 

1. For the $V$ basis  the mean expected value of $k$ scaled like $1.2+ 0.36\, \log_5(1/\varepsilon)$, 
while maximum number of candidates scaled like $8.8+4 \, \log_5(1/\varepsilon)$.

2. Somewhat surprisingly at compilation stage 2 over either Clifford$+T$ basis or Clifford$+\pi/12$ basis $k$ shows very little correlation with the target precision (insignificant correlation coefficient) with mean expectation of $k$ around $2.2$ for the Clifford$+T$ basis and around $2.1$ for the Clifford$+\pi/12$ basis. The expected maximum $k$ also appears uncorrelated and stands at $23$ for the Clifford$+T$ basis, $22$ for the Clifford$+\pi/12$ basis.

Thus, while $k$ had been in single digits for the majority of $(\theta,\varepsilon)$ pairs it occasionally turned up quite high in outlying cases. 
(Notwithstanding, even the outlying cases were classically manageable and finished in seconds due to throttling of integer factorization.)
A conceptual explanation of the apparent stochastic behavior remains to be found. It might be related to the apparent fractal structure of the set of solvable norm equations.

To summarize, we find that the runtime expectation of all three flavors of the algorithm is linear in $\log(1/\varepsilon)$ at coarse precisions, and becomes quadratic in $\log(1/\varepsilon)$  at finer precisions. Runtime can occasionally spike for outlying $(\theta,\varepsilon)$ pairs due to fluctuations in the required number of modifier candidates, while still being in seconds when  run on a common desktop computer.

We believe the runtime performance can be further optimized by reimplementing the algorithm in a fully-compiled language.






\end{document}